\documentstyle[11pt,epsf,amssymb,xypic]{conm-p-l} 

\expandafter\ifx\csname amsart.sty\endcsname\relax
\expandafter\ifx\csname @ptsize\endcsname\relax
	\stop
\fi
\fi

\expandafter\ifx\csname epsfbox\endcsname\relax
	\def\epsfbox#1{Insert PostScript(TM) file {\tt#1} here}
\fi

\expandafter\ifx\csname currentvolume\endcsname\relax
\else
	\advance\topmargin by1cm
	\advance\oddsidemargin by.6cm
	\evensidemargin \oddsidemargin
	
	\makeatletter
	\let\serieslogo@\relax
	\setbox\copyrightbox@\hbox{}
	\makeatother
\fi

\makeatletter

\@ifundefined{selectfont}{%
	\let\text\hbox
	\@input{amssym.def}
	\@input{amssym}
}{%
	\input amsfonts.sty
	\input amstext.sty
	\input amssymb.sty
}

\expandafter\ifx\csname theorembodyfont\endcsname\relax

\let\old@newtheorem\newtheorem
\def\newtheorem{\@ifstar\@nt@thmlike\old@newtheorem}
\def\@nt@thmlike#1{%
	\expandafter\let\csname diffthm@#1\endcsname\empty
	\old@newtheorem{#1}}

\def\thmstylei{\rm}
\def\thmstyleii{\em}
\def\@ifdiffthm#1#2{\xdef\@diffthm{diffthm@\@currenvir}%
	\expandafter\ifx\csname\@diffthm\endcsname\relax#2\else#1\fi}
\def\indentedtrivlist{\parsep\parskip
  \@trivlist \labelwidth\z@
  \leftmargin\leftmargini  \leftskip\leftmargini
  \expandafter\ifx\csname mathindent\endcsname\relax\else
	\multiply\mathindent by 2
  \fi
  \itemindent\z@ \def\makelabel##1{##1}}

\let\thmlisti\trivlist
\let\thmlistii\trivlist
\let\thmopargstyle\bf

\def\@begintheorem#1#2{%
	\@ifdiffthm{
		\thmlistii \item[\ifx\thmlisti\indentedtrivlist\hskip-\leftmargini\fi
			\hskip\labelsep{\bf #1\ #2:}]
		\thmstyleii\relax
		\ifx\thmlistii\indentedtrivlist\strut\\\fi
	}{
		\thmlisti \item[\ifx\thmlisti\indentedtrivlist\hskip-\leftmargini\fi
			\hskip\labelsep{\bf #1\ #2:}]
		\thmstylei \relax
		\ifx\thmlisti\indentedtrivlist\strut\\\fi
	}%
	\relax}
\def\@opargbegintheorem#1#2#3{%
	\@ifdiffthm{
		\thmlistii \item[\ifx\thmlisti\indentedtrivlist\hskip-\leftmargini\fi
			\hskip\labelsep{\bf #1\ #2\  {\thmopargstyle(#3)}:}]
		\thmstyleii\relax
		\ifx\thmlisti\indentedtrivlist\strut\\\fi
	}{
		\thmlisti \item[\ifx\thmlisti\indentedtrivlist\hskip-\leftmargini\fi
			\hskip\labelsep{\bf #1\ #2\  {\thmopargstyle(#3)}:}]
		\thmstylei \relax
		\ifx\thmlisti\indentedtrivlist\strut\\\fi
	}%
	\index{#3}\relax}

\fi 

\def\latexfmtname{lplain}
\ifx\fmtname\latexfmtname
\@ifundefined{NoWithin}{%
\@ifundefined{chapter}{%
	\@ifundefined{theorembodyfont}{
		\newtheorem*{theorem}{Theorem}[section]
	}{
		\newtheorem{theorem}{Theorem}[section]
	}
	\newtheorem{algorithm}{Algorithm}[section]
	\newtheorem{definition}{Definition}[section]
	\newtheorem{example}{Example}[section]
}{
	\@ifundefined{theorembodyfont}{
		\newtheorem*{theorem}{Theorem}[chapter]
	}{
		\newtheorem{theorem}{Theorem}[chapter]
	}
	\newtheorem{algorithm}{Algorithm}[chapter]
	\newtheorem{definition}{Definition}[chapter]
	\newtheorem{example}{Example}[chapter]
}  
}{
	\@ifundefined{theorembodyfont}{
		\newtheorem*{theorem}{Theorem}
	}{
		\newtheorem{theorem}{Theorem}
	}
	\newtheorem{algorithm}{Algorithm}
	\newtheorem{definition}{Definition}
	\newtheorem{example}{Example}
}  

\@ifundefined{theorembodyfont}{
	\newtheorem*{lemma}[theorem]{Lemma}
	\newtheorem*{corollary}[theorem]{Corollary}
	\newtheorem*{proposition}[theorem]{Proposition}
}{
	\newtheorem{lemma}[theorem]{Lemma}
	\newtheorem{corollary}[theorem]{Corollary}
	\newtheorem{proposition}[theorem]{Proposition}
}
\newtheorem{exercise}[example]{Exercise}
\newtheorem{actualexamples}[example]{Examples}
\newtheorem{actualproblems}[example]{Problems}
\newtheorem{definitions}[definition]{Definitions}
\fi

\def\atendofproof{{\ifvmode\indent\fi
	\unskip\nobreak\hfil\penalty50\vadjust{\penalty500}%
	\hskip2em\hbox{}\nobreak\hfil$\Box$
	\parfillskip=0pt\finalhyphendemerits=0\par}}
\@ifundefined{qed}{\let\qed=\atendofproof 
	}{\let\atendofproof\qed 
	}

\def\listnopb{\let\@beginparpenalty=\@M}

\@ifundefined{theorem@headerfont}{
	\def\titledparfont{\bf}
}{
	\let\titledparfont\theorem@headerfont
}
\@ifundefined{theorem@indent}{
	\def\titledparindent{\noindent}
}{
	\let\titledparindent\theorem@indent
}
\long\def\titledpar#1{\par\titledparindent{\titledparfont #1}}

\newenvironment{remark}{\titledpar{Remark.}}{\par}
\newenvironment{remarks}{\titledpar{Remarks.}}{\par}

\def\proofsec{{\def\check{theorem}%
	\ifx\@currentreference\check\else
		\par\medskip\rm\fi}
	\@ifnextchar[{\@proofsecarg}{\@proofsecnoarg}}
\def\@proofsecarg[#1]{\titledpar{Proof #1. }}
\def\@proofsecnoarg{\titledpar{Proof. }}

\@ifundefined{pf}{
	\newenvironment{proof}{\ifvmode \else\par\fi\nobreak\smallskip\nobreak
		\proofsec}{\atendofproof\medbreak}
}{
	\newenvironment{proof}{\pf}{\endpf}
}

\@namedef{examples*}{\begin{actualexamples}}
\@namedef{endexamples*}{\end{actualexamples}}
\newenvironment{olddefinitions}{\ifvmode \else\par\fi\medbreak
	\noindent{\bf Definitions:}\par\nobreak
	\begin{itemize}\listnopb}{\end{itemize}\par}
\def\OldDefinitions{\let\definitions\olddefinitions
	\let\enddefinitions\endolddefinitions}

\newcounter{problem}
\newcounter{examplectr}
\@addtoreset{problem}{example}
\@addtoreset{examplectr}{example}

\def\labelproblem{{\bf\arabic{problem}.}}
\def\labelexamplectr{{\bf(\roman{examplectr})}}
\newenvironment{problems}{\@ifnextchar[\problemsoptarg{\problemsoptarg[]}%
}{\endlist\end{actualproblems}}
\def\problemsoptarg[#1]{\begin{actualproblems}%
	{\def\check{c}\def\checkii{#1}%
		\ifx\check\checkii (continued)\else
		\def\check{}\ifx\check\checkii \else(\checkii)\fi\fi}%
	\strut\par\nobreak\global\@nobreaktrue
	\list{\labelproblem}{\usecounter{problem}%
		\def\makelabel##1{\hss\llap{##1}}}\relax}
\newenvironment{examples}{\goodbreak
	\begin{actualexamples}%
	\global\@nobreaktrue
	\strut\par\nobreak\global\@nobreaktrue
	\list{\labelexamplectr}{\usecounter{examplectr}%
		\def\makelabel##1{\hss\llap{##1}}}\relax
	\nobreak\global\@nobreaktrue
}{\endlist\end{actualexamples}}

\@ifundefined{selectfont}{
	\let\backtoroman\rm
}{
	\let\backtoroman\reset@font
}
\def\bal{\begin{algorithm}\backtoroman}		\def\eal{\end{algorithm}}
\def\ba{\begin{array}}			\def\ea{\end{array}}
\def\bc{\begin{corollary}}		\def\ec{\end{corollary}}
\def\bde{\begin{description}}		\def\ede{\end{description}}
\def\bds{\begin{definitions}\backtoroman}		\def\eds{\end{definitions}}
\def\bd{\begin{definition}\backtoroman}		\def\ed{\end{definition}}
\def\ben{\begin{enumerate}}		\def\een{\end{enumerate}}
\def\be{\begin{eqnarray*}}		\def\ee{\end{eqnarray*}}
\def\bfig{\begin{figure}}		\def\efig{\end{figure}}
\def\bi{\begin{itemize}}		\def\ei{\end{itemize}}
\def\bl{\begin{lemma}}			\def\el{\end{lemma}}
\def\bne{\begin{eqnarray}}		\def\ene{\end{eqnarray}}
\def\bp{\begin{proof}}			\def\ep{\end{proof}}
\def\bse{\begin{equation}}		\def\ese{\end{equation}}
\def\btabb{\begin{tabbing}}		\def\etabb{\end{tabbing}}
\def\btable{\begin{table}}		\def\etable{\end{table}}
\def\btab{\begin{tabular}}		\def\etab{\end{tabular}}
\def\bcen{\begin{center}}		\def\ecen{\end{center}}
\def\bt{\begin{theorem}}		\def\et{\end{theorem}}
\def\bpr{\begin{proposition}}		\def\epr{\end{proposition}}
\def\bv{\begin{verbatim}} 		\def\ev{\end{verbatim}}
\def\bxe{\begin{exercise}\backtoroman}		\def\exe{\end{exercise}}
\def\bxs{\begin{examples}\backtoroman}		\def\exs{\end{examples}}
\def\bx{\begin{example}\backtoroman}		\def\ex{\end{example}}
\def\bprob{\begin{problems}\backtoroman}		\def\eprob{\end{problems}}

\def\tf{\textstyle\frac}

\@ifundefined{selectfont}{%
	\def\mydefloglike#1#2#3{\expandafter\def\csname#1\endcsname{\mathop{\rm#2}#3}}
	\let\operator@font\rm
}{%
	\def\mydefloglike#1#2#3{\expandafter\def\csname#1\endcsname
		{\mathop{\operator@font #2}#3}}
}

	\def\half{{\tf12}}

	\newbox\@defeqbox \setbox\@defeqbox=\hbox{\footnotesize def}
	\def\defeq{\mathop{=}\limits^{\copy\@defeqbox}}
	\def\nin{\not\in}
	\def\spdot{\,\cdot\,}
	\def\seq#1#2{\ifmmode \{#1_{#2}\}\else $\{#1_{#2}\}$\fi}
	\let\sub=\subset
	\def\@psup{\mathop{\vtop{\ialign{##\crcr
		$\m@th\supset$\crcr
		\hfil$\m@th\scriptstyle\not=$\crcr}}}}
	\@ifundefined{supsetneqq}{\let\psup\@psup}{\let\psup=\supsetneqq}
	\def\@psub{\mathop{\vtop{\ialign{##\crcr
		$\m@th\subset$\crcr
		\hfil$\m@th\scriptstyle\not=$\crcr}}}}
	\@ifundefined{subsetneqq}{\let\psub\@psub}{\let\psub=\subsetneqq}
	\@ifundefined{preccurlyeq}{}{}
	\@ifundefined{npreceq}{}{}
	\mydefloglike{sgn}{sgn}\nolimits 

	\let\oldlor=\lor
	\let\oldland=\land
	\def\lor{\mathrel\oldlor}
	\def\land{\mathrel\oldland}

	\def\implies{\,\Longrightarrow\,}

	\def\inv#1{{#1}^{-1}}
	\def\set#1for#2\eset{\left\{\:#1\mathbin\setmidchar#2\:\right\}}
	\let\setmidchar:
	
	\mydefloglike{supp}{supp}\nolimits 

	\let\sm\setminus \let\es\emptyset
	\def\lp{\ifmmode\ell^p\else$\ell^p$\fi}

	\def\net#1#2#3{\ifmmode\set#1_{#2} for #2\in#3\eset
		\else$\net{#1}{#2}{#3}$\fi}

	\mydefloglike{relint}{relint}\nolimits
	\mydefloglike{co}{co}\nolimits 
	\mydefloglike{clco}{\overline{co}}\nolimits 
	\mydefloglike{diam}{diam}\nolimits 
	\mydefloglike{bdy}{bdy}\nolimits

	\mydefloglike{cor}{cor}\nolimits 
	
	\mydefloglike{Im}{Im}\nolimits 
	\mydefloglike{im}{im}\nolimits 
	\mydefloglike{Ker}{Ker}\nolimits 
	\mydefloglike{ker}{ker}\nolimits 
	\mydefloglike{rank}{rank}\nolimits 
	\mydefloglike{nullity}{nullity}\nolimits 
	\mydefloglike{ch}{ch}\nolimits 
	\mydefloglike{diag}{diag}\nolimits 
	\mydefloglike{lcm}{lcm}\nolimits 
	\mydefloglike{lin}{lin}\nolimits 
	\mydefloglike{tr}{tr}\nolimits 
	\mydefloglike{sp}{sp}\nolimits 
	\mydefloglike{adj}{adj}\nolimits 

	\mydefloglike{gp}{gp}\nolimits 
	\mydefloglike{id}{id}\nolimits 
	\mydefloglike{charac}{char}\nolimits 
	\mydefloglike{Coker}{Coker}\nolimits 
	\mydefloglike{Coim}{Coim}\nolimits 
	\mydefloglike{Ext}{Ext}\nolimits 
	\mydefloglike{Tor}{Tor}\nolimits 
	\mydefloglike{Hom}{Hom}\nolimits 
	\mydefloglike{End}{End}\nolimits 
	\mydefloglike{Inn}{Inn}\nolimits 
	\mydefloglike{Aut}{Aut}\nolimits 
	\mydefloglike{GF}{GF}\nolimits 
	\mydefloglike{charac}{char}\nolimits 
	\mydefloglike{Spec}{Spec}\nolimits 
	\mydefloglike{Max}{Max}\nolimits 
	\mydefloglike{Rad}{Rad}\nolimits 
	
	\def\pmod#1{\allowbreak
		\ifinner\mkern9mu\else\mkern18mu\fi({\hbox{mod}}\,\,#1)}
	\mydefloglike{STS}{STS}\nolimits 
	\mydefloglike{TTS}{TTS}\nolimits 
	\mydefloglike{DTS}{DTS}\nolimits 
	\mydefloglike{BIBD}{BIBD}\nolimits 
	\mydefloglike{SBIBD}{SBIBD}\nolimits
	\mydefloglike{grad}{grad}\nolimits 
	
	\mydefloglike{div}{div}\nolimits 
	\mydefloglike{curl}{curl}\nolimits 
	\mydefloglike{Int}{Int}\nolimits 
	\mydefloglike{inte}{int}\nolimits 
	\mydefloglike{erf}{erf}\nolimits 
	\mydefloglike{erfc}{erfc}\nolimits 
	\mydefloglike{rel}{rel}\nolimits 
	
	\mydefloglike{GL}{GL}\nolimits
	\mydefloglike{SL}{SL}\nolimits
	\mydefloglike{Sp}{Sp}\nolimits
	\mydefloglike{gl}{gl}\nolimits
	\let\@oldsl\sl
	\def\@newsl{\ifmmode\mathop{\operator@font sl}\nolimits\else\@oldsl\fi}
	\def\sl{\protect\@newsl}
	\mydefloglike{sp}{sp}\nolimits
	\def\o{^\bgroup\c@rc}
	\def\c@rc{\circ\futurelet\next\c@rcs}
	\def\c@rcs{\ifx*\next\let\nxt\c@rc@s \else\ifx^\next\let\nxt\c@rc@t
		\else\let\nxt\egroup\fi\fi \nxt}
	\def\c@rc@s#1{\c@rc} \def\c@rc@t#1#2{#2\egroup}

	\def\DiracNotation{
		\mathcode`\<="8000 {\catcode`\<=\active \gdef<{{\delimiter"426830A}}}
		\mathcode`\>="8000 {\catcode`\>=\active \gdef>{{\delimiter"526930B}}}
		\mathcode`\@="8000 {\catcode`\@=\active \gdef@{{\dagger}}}
	}
		
	\mydefloglike{Re}{Re}\nolimits 
	\mydefloglike{Im}{Im}\nolimits 
	\mydefloglike{Arg}{Arg}\nolimits 
	\mydefloglike{Log}{Log}\nolimits 
	\mydefloglike{res}{res}\nolimits 
	\mydefloglike{ord}{ord}\nolimits 
	\mydefloglike{arg}{arg}\nolimits
	\mydefloglike{cosec}{cosec}\nolimits
	\mydefloglike{edge}{edge}\nolimits 
	\mydefloglike{Lor}{Lor}\nolimits 

\let\overbar\overline

\def\Mhat{{\widehat{\cM}}}

\def\({\left(}
\def\){\right)}
\def\[{\left[}
\def\]{\right]}

\newcount\whichmat \whichmat=0
\newenvironment{pmat}{\whichmat=1 
	\left(\begin{array}}{\end{array}\right)}
\newenvironment{bmat}{\whichmat=2 
	\left[\begin{array}}{\end{array}\right]}
\newenvironment{vmat}{\whichmat=3 
	\left\vert\begin{array}}{\end{array}\right\vert}

\def\m#1#2{
	\ifx(#1 \begin{pmat}{#2}	\else
		\ifx[#1 \begin{bmat}{#2}	\else
			\ifx|#1 \begin{vmat}{#2}	\else
				\begin{#1mat}{#2}
	\fi	\fi	\fi
}

\def\me{
	\ifcase\whichmat
		\errmessage{Error in use of Chris' macros:
			\backslash me when not in a
			matrix environment.}\or
		\end{pmat}\or
		\end{bmat}\or
		\end{vmat}\or
		\end{Vmat}\or
		\end{emat}\or
		\end{cmat}\or
		\end{gmat}\or
		\end{hmat}\or
		\end{qmat}
	\fi
	\whichmat=0
}

\def\thisenumparens#1{%
	\@namedef{the\@enumctr}{\rm(\csname#1\endcsname{\@enumctr})}%
	\@namedef{label\@enumctr}{\rm\csname the\@enumctr\endcsname}%
	}
\def\enumparens#1#2{%
	\@namedef{theenum#1}{\rm(\csname #2\endcsname{enum#1})}%
	\@namedef{labelenum#1}{\rm\csname theenum#1\endcsname}%
	}
\def\thisenumparen#1{%
	\@namedef{the\@enumctr}{\rm\csname #1\endcsname{\@enumctr}}%
	\@namedef{label\@enumctr}{\rm\csname the\@enumctr\endcsname)}%
	}
\def\enumparen#1#2{%
	\@namedef{theenum#1}{\rm\csname #2\endcsname{enum#1}}%
	\@namedef{labelenum#1}{\rm\csname theenum#1\endcsname)}%
	}
\def\thisenumdot#1{%
	\@namedef{the\@enumctr}{\rm\csname #1\endcsname{\@enumctr}}%
	\@namedef{label\@enumctr}{\rm\csname the\@enumctr\endcsname.}%
	}
\def\enumdot#1#2{%
	\@namedef{theenum#1}{\rm\csname #2\endcsname{enum#1}}%
	\@namedef{labelenum#1}{\rm\csname theenum#1\endcsname.}%
	}
\enumdot i{arabic}
\enumparens{ii}{roman}
\enumparen{iii}{alph}

\def\enumspecial#1{
	\@namedef{the\@enumctr}{{\bf(#1{\arabic{\@enumctr}})}}
	\expandafter\ifx\csname textshade\endcsname\relax
		\@namedef{label\@enumctr}{\csname the\@enumctr\endcsname}
	\else
		\@namedef{label\@enumctr}{%
			\textshade{roundcorners}{\csname the\@enumctr\endcsname}}
	\fi
	\let\old@item\item
	\def\item{\old@item\edef\my@@label{{#1\arabic{\@enumctr}}}%
		\expandafter\enlabel\my@@label}
}

\def\p@enumi{}
\def\p@enumii{}
\def\p@enumiii{}
\def\p@enumiv{}

\def\figlabel#1{\label{fig:#1}}
\def\figref#1{\ref{fig:#1}}

\def\thmlabel#1{\label{thm:#1}}
\def\thmref#1{\ref{thm:#1}}
\def\seclabel#1{\label{sec:#1}}
\def\secref#1{\ref{sec:#1}}
\def\enlabel#1{\label{en:#1}}
\def\enref#1{\ref{en:#1}}
\def\eqlabel#1{\label{eq:#1}}
\def\eqref#1{(\ref{eq:#1})}

\def\underuparrow#1{\mathop{\vtop{\ialign{##\crcr
  $\hfil\displaystyle\strut{#1}\hfil$\crcr\noalign{\kern3\p@\nointerlineskip}
  $\hfil\uparrow\hfil$\crcr\noalign{\kern3\p@}}}}\limits}

\def\underaccent#1#2{\smash{%
	\vtop{\ialign{\hfil##\hfil\crcr
	$\m@th#2{}$\crcr\noalign{\kern2pt\nointerlineskip}%
	\vtop to1pt{\hbox{$\m@th#1{\phantom{x}}$}\vss}\crcr
	\noalign{\kern1pt\nointerlineskip}}}}}

\def\circover#1{\vbox{\ialign{\hfil##\hfil\crcr
	$\m@th\smash{\scriptscriptstyle\circ}$\crcr\noalign{\kern1pt\nointerlineskip}
	$\m@th#1$\crcr}}}

\@ifundefined{selectfont}{
	\def\bold#1{{\Bbb#1}}
	\let\vec=\bold
}{
	\def\vec#1{{\bold{#1}}}
}

\mathcode`\^^V="8000
{\catcode`\^^V=\active\global\def^^V{\vec}} 

	\def\eq{&=&}

	\def\age{&\ge&}
	\def\agt{&>&}
	\def\ale{&\le&}
	\def\alt{&<&}

\@ifundefined{selectfont}{%
	\def\range#1#2#3{\hbox{${#1}_{#2},\ldots,{#1}_{#3}$}}
}{%
	\def\range#1#2#3{\hbox{$#1_{#2},\ldots,#1_{#3}$}}
}

\@ifundefined{cL}{}{\message{XY-Pic must be modified,
	see Chris.}\stop}
\def\cA{{\cal A}} \def\cB{{\cal B}}  
\def\cE{{\cal E}}   
\def\cM{{\cal M}} \def\cN{{\cal N}}

\def\cU{{\cal U}} \def\cV{{\cal V}} \def\cW{{\cal W}} 

\let\p=\rho
\let\vp=\varphi  
\let\e=\epsilon

\let\d=\partial 

\let\g=\gamma

\let\l=\lambda

\let\a=\alpha

\def\o{o}
\let\W=\Omega

\let\s\sigma
\let\th=\theta

\expandafter\ifx\csname Bbb\endcsname\relax 
	\def\Bbb#1{{\bf#1}}

\else 
	\expandafter\ifx\csname selectfont\endcsname\relax
	\else
	\fi
\fi
\@ifundefined{RSScript}{%
	\@ifundefined{EuScript}{%

	}{%
		
	}%
	
}{%
	\@ifundefined{EuScript}{%
		
	}{}%
	
}

\def\R{\@ifnextchar[\@Roptarg\@Rnooptarg}
\def\@Roptarg[#1]{{\Bbb R^{#1}}}
\def\@Rnooptarg{{\Bbb R}}
	\let\R\@Rnooptarg
\expandafter\ifx\csname C\endcsname\relax 
	\def\C{{\Bbb C}}
\fi
\def\N{{\Bbb N}}

\def\Z{{\Bbb Z}}

\newdimen\maxpspicwid \maxpspicwid=\textwidth
\def\epsfsize#1#2{\ifdim#1<\maxpspicwid#1\else\maxpspicwid\fi}

\mathcode`\*="8000
{\catcode`*=\active \gdef*{^\bgroup\st@r}}
\def\st@r{\ast\futurelet\next\st@rs}
\def\st@rs{\ifx*\next\let\nxt\st@@@s \else\ifx^\next\let\nxt\st@@@t
	\else\let\nxt\egroup\fi\fi \nxt}
\def\st@@@s#1{\st@r} \def\st@@@t#1#2{#2\egroup}

\newdimen\@qbwid
\newdimen\@qbpari\@qbpari=\parindent
\def\quotebox#1{\@qbwid=\hsize\advance\@qbwid by-#1\@qbpari
	\parbox{\@qbwid}}

\newif\iffirstslide \firstslidetrue
\@ifundefined{@notinslitex}{}{
	\def\newslide{\iffirstslide\firstslidefalse\else\end{slide}\fi
		\@ifstar{\firstslidetrue}{\begin{slide}{}}}
	\def\section#1{\medbreak\noindent{\bf #1}\medbreak}
	\let\subsection\section
	\let\subsubsection\section
	\let\paragraph\section
	\let\chapter\section
}

\@ifundefined{citeasnoun}{%
	\let\citeasnoun\cite
}{%
}

\def\SmallDiagram{%
	\let\objectstyle\scriptstyle
	\spreaddiagramrows{-.5pc}%
	\spreaddiagramcolumns{-.5pc}%
}

\def\SpecialsFor#1 {%
	\ifnum#1=272\relax
		\mathcode`\*="2203 
		\def\D##1##2{\hbox{$D{##1}({##2})$}} 
		\def\C##1{\hbox{$C^{##1}$}} 
		\def\covspacen##1##2{\cE^{##1\ast}_{##2}} 
		\def\covspace{\covspacen n}	
		\let\R\@Rnooptarg 
	\fi
}

\expandafter\ifx\csname selectfont\endcsname\relax
	\def\sfmath#1{{\sf#1}}
	\def\sfslmath#1{{\sf#1}}
\else
	\newmathalphabet\sfmath
	\newmathalphabet*\sfslmath{cmss}{m}{sl}
	\addtoversion{normal}\sfmath{cmss}{m}{n}
	\addtoversion{bold}\sfmath{cmss}{bx}{n}
\fi

\expandafter\ifx\csname amstex.sty\endcsname\relax\else
	
	\def\endequation{\eqno \hbox{\@eqnnum}
	$$\global\@ignoretrue}
	\def\cases#1{\left\{\,\vcenter{\normalbaselines\m@th
		\ialign{$##\hfil$&\quad##\hfil\crcr#1\crcr}}\right.}
\fi

\@ifundefined{rom}{%
	\@ifundefined{selectfont}{%
		\def\rom#1{{\rm#1}}
	}{%
		\def\rom#1{{\shape{n}\selectfont#1}}
	}
}

\expandafter\ifx\csname LeavePageSizes\endcsname\relax
	\evensidemargin\oddsidemargin
\fi

\makeatother

\expandafter\ifx\csname amstex.sty\endcsname\relax

\title{%
	Invariance Properties of Boundary Sets of Open Embeddings of
	Manifolds and Their Application to the Abstract Boundary%
	\protect\footnotetext{
		{\em 1991 MR subject classifications\/}:
		Primary: 53C23, 57R40;
		Secondary: 57R15, 83C75.
	}
}
\author{
	Christopher J. Fama%
	\thanks{%
		Research supported by
		an Australian Postgraduate Award.
		Email address: {\tt Chris.Fama@anu.edu.au}.
		Physical address: Centre for Mathematics and its Applications,
		School of Mathematical Sciences,
		The Australian National University,
		Canberra 0200, Australia.
	}
	\and
	Susan M. Scott%
	\thanks{%
		Research supported by
		an ARC Australian Research Fellowship,
		an ARC Large Research Grant \#$\,$A69332184,
		and
		an ARC Small Research Grant \#$\,$92FIGR076,
		It was also supported in part
		by the National Science Foundation under
		Grant No.~PHY89-04035.
		Email address: {\tt Susan.Scott@anu.edu.au}.
		Physical address: as for CJF.
	}
}
\date{25 March 1994; C.M.A.~preprint \# MRR-036-94, {\tt
gr-qc/9406023}}

\else

\title[INVARIANCE PROPERTIES OF BOUNDARY SETS]{%
	Invariance Properties of Boundary Sets of Open Embeddings of
	Manifolds and Their Application to the Abstract Boundary}
\author{
	Christopher J. Fama
	\and
	Susan M. Scott
	}
\thanks{%
	This research was supported by
	an ARC Australian Research Fellowship (SMS),
	an ARC Large Research Grant \#$\,$A69332184 (SMS),
	an ARC Small Research Grant \#$\,$92FIGR076 (SMS),
	and
	an Australian Postgraduate Award (CJF).
	It was also supported in part (SMS) by the National Science Foundation under
	Grant No.~PHY89-04035.
}
\date{25 March 1994; C.M.A.~preprint \# MRR-036-94, {\tt
gr-qc/9406023}}
\address{Centre for Mathematics and its Applications,
	School of Mathematical Sciences,
	The Australian National University,
	Canberra 0200, Australia.}
\email{Chris.Fama@@anu.edu.au}
\address{Centre for Mathematics and its Applications,
	School of Mathematical Sciences,
	The Australian National University,
	Canberra 0200, Australia.}
\email{Susan.Scott@@anu.edu.au}

\theoremstyle{definition}
\numberwithin{equation}{section}

\expandafter\ifx\csname currentvolume\endcsname\relax \else
\makeatletter
\def\abstract{%
	\global\let\abstract\relax
	\defaultfont\normalsize  \skip@28\p@ \advance\skip@-\lastskip
	\advance\skip@-\baselineskip \vskip\skip@
	\vtop \bgroup
	\noindent
	{\sc\abstractname}.\ \ \ignorespaces
}
\def\endabstract{\strut\par\egroup\smallskip}

\def\@maketitle{%
  \defaultfont\normalsize
  \let\@makefnmark\relax  \let\@thefnmark\relax
  \ifx\@empty\@subjclass\else
   \@footnotetext{1991 {\it Mathematics Subject
     Classification}.\enspace
        \@subjclass.}\fi
  \ifx\@empty\@keywords\else
   \@footnotetext{{\it Key words and phrases.}\enspace \@keywords.}\fi
\ifx\@empty\@thanks\else
   \@footnotetext{\@thanks}\fi
  \topskip 10pc minus 9pc 
  \advance\topskip-\headsep \advance\topskip-\headheight
  \vtop{\centering{\Large\bf\@title\@@par}%
   \global\dimen@i\prevdepth}%
  \prevdepth\dimen@i
  \ifx\@empty\@authors
  \else
    \baselineskip24\p@
    \vtop{\@andify{ AND }\@authors
      \centering{\expandafter\uppercasetext@\expandafter
       {\@authors}\@@par}%
         \global\dimen@i\prevdepth}\relax
    \prevdepth\dimen@i
  \fi
  \ifx\@empty\@dedicatory
  \else
    \baselineskip18\p@
  \vtop{\centering{\small\it\@dedicatory\@@par}%
      \global\dimen@i\prevdepth}\prevdepth\dimen@i
  \fi
  \ifx\@empty\@date\else
  \baselineskip24\p@
    \vtop{\centering\@date\@@par
      \global\dimen@i\prevdepth}\prevdepth\dimen@i
  \fi
  \normalsize
  \vskip32\p@\@plus14\p@
  } 
\makeatletter

\fi
\fi

\expandafter\ifx\csname selectfont\endcsname\relax
\def\TNP{{\rm $T$NP}}
\def\CNP{{\rm $C$NP}}
\def\SCNP{{\rm $SC$NP}}
\def\PIkNP{$\Pi_k${\rm NP}}
\def\HOMkNP{$HOM_k${\rm NP}}
\else
\def\TNP{{\shape{n}\selectfont $T$NP}}
\def\CNP{{\shape{n}\selectfont $C$NP}}
\def\SCNP{{\shape{n}\selectfont $SC$NP}}
\def\PIkNP{$\Pi_k${\shape{n}\selectfont NP}}
\def\HOMkNP{$HOM_k${\shape{n}\selectfont NP}}
\fi

\begin{document}

\maketitle

\maxpspicwid=\hsize


\begin{abstract}
	The {\em abstract boundary\/} (or {\em {\em a\/}-boundary\/}) of Scott and
	Szekeres \cite{Scott94} constitutes a ``boundary'' to any
	$n$-dimensional,
	paracompact, connected, Hausdorff, $C^\infty$-manifold (without a
	boundary in the usual sense).
	In general relativity one deals with a {\em space-time\/} $(\cM,g)$
	(a 4-dimensional manifold $\cM$ with a Lorentzian metric $g$),
	together with a chosen preferred class of curves in $\cM$.
	In this case the
	{\em a\/}-boundary
	points may represent ``singularities'' or ``points at infinity''.
	Since the
	{\em a\/}-boundary
	itself, however, does not depend on the existence of further
	structure on the manifold such as a Lorentzian metric or
	connection, it is possible for it to be
	used in many contexts.

	In this paper we develop some purely topological
	properties of abstract boundary sets and abstract boundary points
	({\em a\/}-boundary points).
	We prove, amongst other things,
	that compactness is invariant under boundary set equivalence,
	and introduce another invariant concept ({\em isolation\/}),
	which encapsulates the notion that a boundary set is ``separated''
	from other boundary points of the same embedding.
	It is demonstrated, however, that the properties of connectedness
	and simple connectedness of boundary sets
	are {\em not\/} invariant under boundary set equivalence.

	We introduce a {\em $T$ neighbourhood property\/} (\TNP) of boundary
	sets, where $T$ is any {\em topological property\/} of a topological
	space, i.e., a property invariant under all homeomorphisms of the
	space. It is shown that any such \TNP{} {\em is\/} invariant under
	boundary set equivalence.
	The connected neighbourhood property (\CNP)
	and the simply connected neighbourhood property (\SCNP)
	are examined in detail.  It is proven, for instance, that any
	boundary set satisfying the \CNP{} must be connected.
	We conclude with a consideration of (isolated) boundary sets which are also
	submanifolds of the enveloping manifold.
	When the above concepts appear, throughout the paper, they are
	related to the abstract boundary (i.e., the set of equivalence classes of
	boundary sets which include a singleton).
\end{abstract}


\section{Introduction}
\seclabel{intro}

We briefly summarise the construction of the {\em a\/}-boundary as presented in
\citeasnoun{Scott94}, and fix our notation.
Only the first part of that paper is needed here, before the
introduction of a pseudo-Riemannian metric and the ensuing classification
of {\em a\/}-boundary points.

All manifolds will be assumed to be smooth (i.e., $C^\infty$),
Hausdorff, connected and paracompact manifolds without boundary,
unless otherwise stated.
We call a $C^\infty$  embedding $\phi:\cM\to\Mhat$, where $\cM$, $\Mhat$ are
manifolds {\em of the same dimension\/} $n$, an {\em envelopment},
and denote this $(\cM,\Mhat,\phi)$ (or sometimes, somewhat loosely,
$\phi$).
The topological boundary of the image $\phi(\cM)$ in $\Mhat$, $\d(\phi(\cM))$,
is denoted $\d_\phi\cM$ or just $\d_\phi$.
A {\em boundary set\/} of an envelopment
is a subset $B\sub\d_\phi$.
The abstract boundary consists of special equivalence classes of
boundary sets, in various envelopments, under a certain equivalence
relation.
This addresses the difficulty one faces in defining space-time ``boundary
points'' in general relativity, for instance, where the topology of the
``boundary point'' is certainly not invariant under diffeomorphisms of the
space-time (i.e., various envelopments).

The equivalence relation used is defined as follows.  Suppose that we are
given boundary sets $B,\,B'$ of two envelopments
$(\cM,\Mhat,\phi)$, $(\cM,\Mhat',\phi')$, respectively. We say that
$B$ {\em covers\/} $B'$, in this paper abbreviated to $B\rhd B'$,
if for every open neighbourhood $\cU$ of $B$ in $\Mhat$, there is an open
neighbourhood $\cU'$ of $B'$ in $\Mhat'$ such that
$$
	\phi\circ(\phi'\inv)(\cU'\cap\phi'(\cM)) \sub \cU.
$$
This encapsulates the idea that ``one cannot approach $B'$ from within
$\cM$ without also approaching $B$''.
In fact, using this notation,  we quote a theorem:

\smallskip
\titledpar{Theorem 19 of \citeasnoun{Scott94}.}
{\em
$B\rhd B'$ if and only if, for {\bf any\/}
sequence \seq pi of points of $\cM$ such that the sequence
$\{\phi'(p_i)\}$ has a limit
point in $B'$, the sequence $\{\phi(p_i)\}$ has a limit point in $B$.
}

\medskip

One then says that boundary sets $B$ and $B'$ are {\em equivalent},
$B\sim B'$, if they cover each other: $B\rhd B'$ and $B\lhd B'$.
The equivalence class containing the boundary set $B$ is denoted $[B]$,
and is called an {\em abstract boundary set}.
If $[B]$ contains a singleton boundary set $\{p\}$, then $[B]$ may
also be denoted by $[p]$, and is called an {\em abstract boundary point\/}.%

The collection of all abstract boundary points constitutes
the {\em abstract boundary\/} $\cB(\cM)$ of $\cM$; that is,
$$
	\cB(\cM) \defeq
	\set [p] for p\in\d_\phi\cM \hbox{ for some envelopment }
		(\cM,\Mhat,\phi) \eset.
$$
For further details of this
construction, see
\citeasnoun{Scott94}.

We will assume that primed objects belong to an envelopment
$(\cM,\Mhat',\phi')$ unless otherwise stated, with corresponding
unprimed objects belonging to
an envelopment $(\cM,\Mhat,\phi)$.

\section{Compact boundary sets}

We note that any paracompact, connected, Hausdorff, smooth manifold $\cM$
can be given a Riemannian metric $g_0$ by using a partition of unity to
``smooth out'' Riemannian metrics on the domains of local coordinate charts
(see, e.g., \citeasnoun{Gallot87}).  Further, any (smooth)
Riemannian metric $g_0$
on $\cM$ is globally conformal to a (smooth) complete Riemannian metric $g$
\cite{Nomizu61}, i.e., $g=\W^2 g_0$ where $\W$ is a smooth,
positive-valued function on $\cM$.
Now $g$ induces a complete [topological] metric $d$,
which is the distance function given by $d(p,q)=\inf\{$Riemannian
arc lengths of piecewise smooth curves
connecting $p$ and $q\}$.%
\footnote{
	Note that the Hopf-Rinow Theorem states that if $g$ is complete then
	the phrase
	``piecewise smooth curves'' may be replaced with ``smooth
	geodesics''.
}
We shall use the term ``Riemannian metric'' for $d$ as well as for $g$,
and we will assume them to be complete, unless otherwise stated.

The metric topology of $d$ agrees
with the given manifold topology on $\cM$.
Since $d$ is complete,
the Hopf-Rinow Theorem (see, e.g., \citeasnoun{Beem81})
says that every subset $K$ of $\cM$ for which $d$ is bounded (i.e.,
$\sup\set d(p,q) for p,q\in K\eset < \infty$)
has compact closure.  The ability to connect compactness and boundedness
in this way is the reason we implement complete metrics extensively
throughout this paper.

	\begin{figure}[p]
		\centering\leavevmode
		\epsfbox{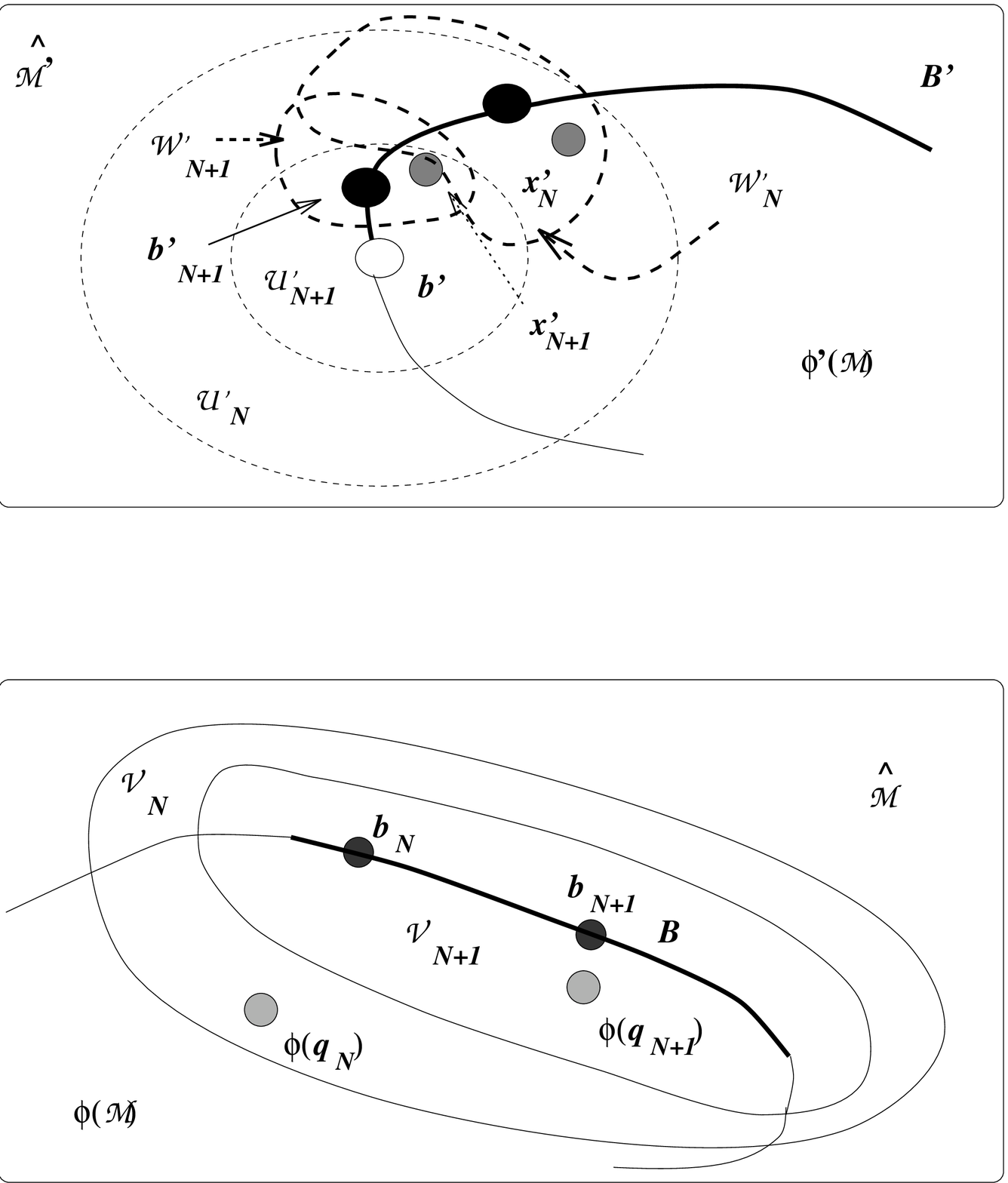}
		\caption{Proof of
			Lemma~\protect\thmref{B compact sim B' imp B' closed}.
			\figlabel{closure lem}}
	\end{figure}
The following lemma demonstrates that if a boundary set of a given
envelopment is compact, then all boundary sets (of other envelopments)
which are equivalent to it must be closed.

\begin{lemma}
	\thmlabel{B compact sim B' imp B' closed}
	If a boundary set $B\sub\d_\phi$ is compact, then any $B'\sim B$ is
	closed.
\end{lemma}
\begin{proof}
	(As usual $B\sub\d_\phi,B'\sub\d_{\phi'}$.)

	Let $d,d'$ be complete Riemannian metrics
	on $\Mhat,\Mhat'$ respectively. Define
	$$
		\cV_N \defeq \set y\in \Mhat for d(y,b) < 1/N \hbox{ for some }
		b\in B \eset, \qquad N\in\N.
	$$
	Now suppose that $B'\sim B$ and that $b'\in\overbar{B'}\sm B'$.
	Define
	$$
		\cU'_N\defeq\set x'\in\Mhat' for d'(b',x') < 1/N \eset,\qquad
		N\in\N.
	$$
	There exists $b_N'\in\cU'_N\cap B'$, for each $N$.

	Since $B\rhd b_N'$ \cite[Theorem~16]{Scott94}, there exists an
	open neighbourhood $\cW'_N$ of $b'_N$ in $\Mhat'$ such that
	$$
		\phi\circ(\phi'\inv)(\cW'_N\cap\phi'(\cM)) \sub \cV_N.
	$$
	Since $b'_N$ is in the open set $\cU'_N\cap \cW'_N$ and
	$b_N'\in\d_{\phi'}$, there clearly exists an
	$x_N'\in \cU'_N\cap \cW'_N \cap\phi'(\cM)$.
	See Figure~\figref{closure lem}.

	Let $q_N\defeq(\phi'\inv)(x'_N)$, $\forall\; N\in\N$, so that \seq qN
	is a sequence of points in $\cM$ such that $\phi'(q_N)=x'_N\to b'$.
	Since $\phi(q_N)\in \cV_N$, there is a $b_N\in B$ such that
	$d(b_N,\phi(q_N))<1/N$.

	Now \seq bN is an infinite sequence of
	boundary points in $B$ and so must have a limit point $b*$ in $B$
	(since $B$ is compact).
	Choose the first element $b_{N_1}$ of the sequence \seq bN that
	satisfies $d(b*,b_{N_1})<1$ and set $p_1=q_{N_1}$.
	Next choose the first element $b_{N_2}$ of the sequence \seq bN occurring
	after $b_{N_1}$ that satisfies $d(b*,b_{N_2})<1/2$ and set $p_2=q_{N_2}$.
	Continue in this manner to form the infinite sequence \seq pi in $\cM$
	where for any $i\in\N$, $d(b*,b_{N_i})<1/i$.
	Since $i\le N_i$, $\forall\; i\in\N$,
	\begin{eqnarray*}
		d(b_{N_i},\phi(p_i))
		\eq d(b_{N_i},\phi(q_{N_i}))\\
		\alt 1/N_i\\
		\ale 1/i,
	\end{eqnarray*}
	and
	\begin{eqnarray*}
		d(b*,\phi(p_i))
		\ale d(b*, b_{N_i}) + d(b_{N_i},\phi(p_i))\\
		\alt 1/i + 1/i = 2/i.
	\end{eqnarray*}
	Thus $\phi(p_i)\to b*$.

	Since $B'\rhd B$, the sequence $\{\phi'(p_i)\}$ must have a
	limit point in $B'$
	\cite[Theorem~19]{Scott94}. However, $\phi'(q_N)\to b'$ and so
	$\phi'(p_i)\to b'\nin B'$, which gives a contradiction.

	Thus $\overbar{B'}=B'$.
\end{proof}

Note that in the above lemma it is {\em not\/} sufficient to assume that
$B$ is merely closed, as opposed to compact.  That is, closedness is not
invariant under boundary set equivalence, as will be clearly demonstrated
by the following example.

\begin{example}\backtoroman
	\label{closed not stable ex}
	Let $\Mhat=\R^3$, $\cM=\set (x,y,z)\in\R^3 for z<0\eset$,
	$\phi={}$inclusion,
	and $B=\d_\phi=\{z=0\}$, a plane (closed, but non-compact).

	Now compactify $\cM$ to an open box $\{|x|,|y|,|z|<1,z<0\}$
	in $\R^3$. So $\Mhat'=\R^3$, $\phi'(\cM)=\{|x|,|y|,|z|<1,z<0\}$, and $B\sim
	B'$ where $B'$ is the ``top face'' $\{z=0,|x|,|y|<1\}$ of $\phi'(\cM)$,
	which does {\em\bf not\/} include the perimeter
	$\{z=0,\,|x|,|y|\le1$ and either $|x|$ or $|y|=1\}$.
	$B'$ is most certainly not closed.
\end{example}

	\begin{figure}[p]
		\centering\leavevmode
		\epsfbox{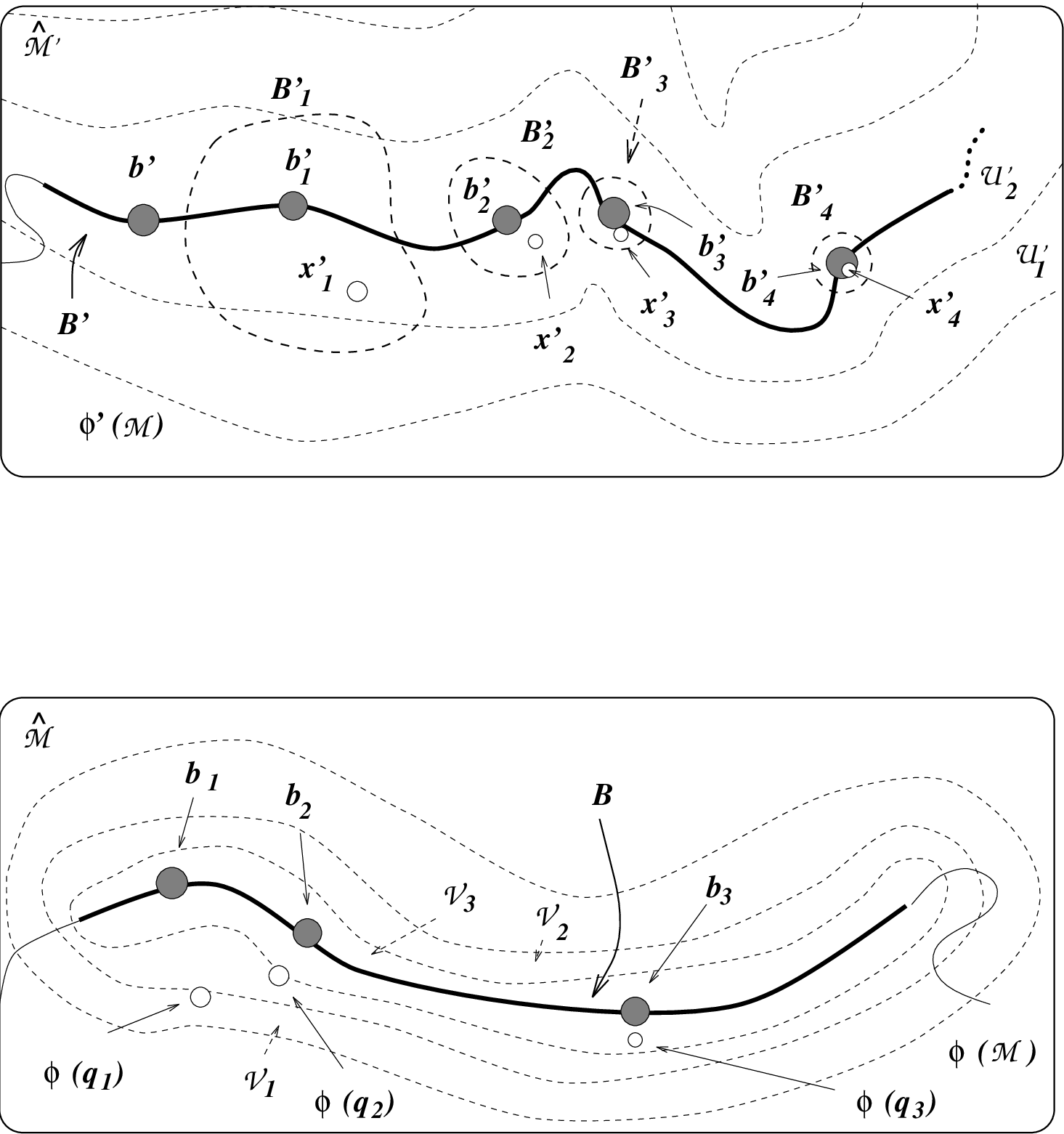}
		\caption{Proof of
			Theorem~\protect\thmref{compactness stable}.
			\figlabel{compact lem}}
	\end{figure}
Theorem~\thmref{compactness stable}
which follows establishes that compactness {\bf\em is\/} invariant under
boundary set equivalence.  So if a boundary set of a given envelopment
is compact, then all boundary sets equivalent to it must be compact.
This, of course, implies that an abstract boundary set $[B]$ may be
labelled ``compact'' if and only if $B$ is compact.

\begin{theorem}
	\thmlabel{compactness stable}
	If a boundary set $B\sub\d_\phi$ is compact, then any $B'\sim B$ is
	compact.
\end{theorem}
\begin{proof}
	Assume that $B'$ is not compact and let
	$d,d'$ be complete Riemannian metrics on $\Mhat,\Mhat'$
	respectively. Since $B'$ is closed by
	Lemma~\thmref{B compact sim B' imp B' closed}, $B'$ cannot be
	bounded with respect to $d'$ (see the remark at the beginning of this
	section).

	Fix $b'\in B'$, and choose $b'_1\in B'$ so that $d'(b',b'_1)>1$.
	Next choose $b'_2\in B'$ with $d'(b',b'_2)>2$ and $d'(b'_1,b'_2)>1$.
	Continue in this manner to form an
	infinite sequence $\seq{b'}n$ of boundary points in $B'$ such that
	$d'(b'_m,b'_n)>1$, $\forall\:m,n\in\N$ ($m\ne n$), and
	$d'(b',b'_n)>n$, $\forall\:n\in\N$ (this is possible due to the
	unbounded nature of $B'$).  Thus
	$d'(b',b'_n) \to\infty$ as $n\to\infty$.
	Now let
	$$
		B_n'\defeq\set x'\in\Mhat' for d'(b'_n,x')<1/2n\eset,\qquad
		n=1,2,\ldots.
	$$
	Each $B'_n$ is open, and clearly $B_m' \cap B_n'=\es$ for all $m\ne n$.
	Now let
	$$
		\cV_n\defeq\set y\in\Mhat for d(b,y)<1/2n\hbox{ for some }
		b\in B\eset,\qquad n=1,2,\ldots,
	$$
	again open sets (this time in $\Mhat$).

	Since $B\rhd B'$, for each $\cV_n$ there is an open set $\cU'_n$ of
	$\Mhat'$ containing $B'$ such that
	$$
		\phi\circ(\phi'\inv)(\cU'_n\cap\phi'(\cM)) \sub \cV_n.
	$$
	Since $b'_n\in\cU'_n\cap B'_n$ which is open, and
	$b_n'\in\d_{\phi'}$, there exists an
	$x'_n\in \cU'_n\cap B'_n\cap\phi'(\cM)$.
	Let $q_n\defeq(\phi'\inv)(x'_n)$, $\forall\; n\in\N$.
	Since $\phi(q_n)\in \cV_n$, there exists a
	$b_n\in B$ such that $d(b_n,\phi(q_n))<1/2n$.
	See Figure~\figref{compact lem}.

	Exactly as in the proof of
	Lemma~\thmref{B compact sim B' imp B' closed},
	\seq bn is an infinite sequence of
	boundary points in $B$ and so must have a limit point $b*$ in $B$
	(since $B$ is compact).
	Choose the first element $b_{n_1}$ of the sequence \seq bn that
	satisfies $d(b*,b_{n_1})<1/2$, and set $p_1=q_{n_1}$.
	Next choose the first element $b_{n_2}$ of the sequence \seq bn occurring
	after $b_{n_1}$ and satisfying $d(b*,b_{n_2})<1/4$, and set $p_2=q_{n_2}$.
	Continue in this manner to form the infinite sequence \seq pi in $\cM$
	where for any $i\in\N$, $d(b*,b_{n_i})<1/2i$.
	Since $i\le n_i$, $\forall\; i\in\N$,
	\begin{eqnarray*}
		d(b_{n_i},\phi(p_i))
		\eq d(b_{n_i},\phi(q_{n_i}))\\
		\alt 1/2n_i\\
		\ale 1/2i,
	\end{eqnarray*}
	and
	\begin{eqnarray*}
		d(b*,\phi(p_i))
		\ale d(b*, b_{n_i}) + d(b_{n_i},\phi(p_i))\\
		\alt 1/2i + 1/2i = 1/i.
	\end{eqnarray*}
	Thus $\phi(p_i)\to b*$.

	Since $B'\rhd B$, the sequence $\{\phi'(p_i)\}$ must have a
	limit point $\tilde b'$ in $B'$
	\cite[Theorem~19]{Scott94}, and
	$d'(b',\tilde b')=X\in\R,\,X\ge0$. Also
	\begin{eqnarray*}
		d'(b',\phi'(p_i))
		\eq d'(b',\phi'(q_{n_i}))\\
		\eq d'(b',x'_{n_i})\\
		\age d'(b',b'_{n_i}) - d'(b'_{n_i},x'_{n_i})\\
		\agt d'(b',b'_{n_i}) - 1/2n_i.
	\end{eqnarray*}
	Thus $d'(b',\phi'(p_i))\to\infty$ as $i\to\infty$.  However,
	\begin{eqnarray*}
		d'(\tilde b',\phi'(p_i))
		\age d'(b',\phi'(p_i)) - d'(b',\tilde b')\\
		\eq d'(b',\phi'(p_i)) - X.
	\end{eqnarray*}
	It follows that
	$d'(\tilde b',\phi'(p_i))\to\infty$ as $i\to\infty$.
	This contradicts the fact that $\tilde b'$
	is a limit point of the sequence $\{\phi'(p_i)\}$.

	Thus $B'$ must be compact.
\end{proof}

\begin{corollary}
	All boundary sets equivalent to a boundary point are compact.
\end{corollary}

In particular, this means that the label ``compact'' may be transferred to
every abstract boundary point $[p]$
contained in the abstract boundary $\cB(\cM)$ of $\cM$,
since for any boundary set $B$ such that $[B]=[p]$, $B$ must be
compact.
One might ask whether any
compact abstract boundary set $[B]$ exists such that $[B]\nin\cB(\cM)$.
The answer to this question is yes: see
Examples~\ref{ex:disconn comp isol abs set not cont pt}
and~\ref{ex:conn comp isol abs set not cont pt}.

\section{Isolated boundary sets}

\subsection{Definition and properties}

It is often of interest to consider boundary sets which are, in some
sense, not close to any other boundary points of the given envelopment.
This notion is encapsulated by the following definition of an isolated
boundary set.

\begin{definition}\backtoroman
	A compact boundary set $B$ is said to be {\em isolated\/} if
	there is an open neighbourhood (in $\Mhat$) $\cU$ of $B$ such that
	$$
		\cU\cap\d_\phi= B.
	$$
\end{definition}

Note that compactness {\em is\/} assumed here.  The following examples
show why closedness and boundedness are both necessary assumptions of the above
definition.

\begin{examples}\backtoroman
\item
	\enlabel{why isol closed}
	{\bf\em What is wrong with $B$ open?}

	Consider $(\cM,\Mhat,\phi)$ to be the same as in
	Example~\ref{closed not stable ex}.
	Let $\cU$ be an open ball cutting the $z=0$ plane and let
	$B=\cU\cap\d_\phi$, a boundary set which we obviously would
	{\bf\em not\/} wish to
	call ``isolated''---it is in no sense ``separated'' from the rest of
	the boundary (c.f.\ Lemma~\thmref{isol comps}).
\item
	\enlabel{why isol bdd}
	{\bf\em What is wrong with $B$ unbounded?}

	Again we consider
	Example~\ref{closed not stable ex}.
	We note that the boundary set $B$
	(as given in Example~\ref{closed not stable ex})
	would be
	isolated if compactness were not required, but $B\sim B'$ and $B'$
	is again of the type given in
	Example~\enref{why isol closed}
	above which we would not want to call
	``isolated''.
\end{examples}

The sense in which
Examples~\enref{why isol closed} and~\enref{why isol bdd}
are undesirable is expressed precisely in the following useful lemma.
First, we recall some topology.

\smallskip
\expandafter\ifx\csname defaultfont\endcsname\relax \else
\begingroup\def\junk{\defaultfont\sc}\ifx\titledparfont\junk
	\titledpar{Mots deja vus.}
\else
	\titledpar{Mots d\'eja vus.}
\fi
\endgroup
\fi
	A subset $Y$ of a topological space $X$ is {\em connected\/} iff it cannot be
	expressed as the disjoint union of two non-empty sets which are open
	in the relative topology of $Y$.
	Equivalently, $Y$ is connected iff
	it contains no proper, non-empty subsets which are both open and
	closed
	in the relative topology of $Y$.
	A subset which is not connected is {\em disconnected}.

	A {\em component\/} of a topological space is a maximal
	connected subset, i.e., a connected subset properly contained in no other.
	A topological space is the disjoint union of its components.

	Note that {\bf we do not require that components be open}
	because then, for instance, the closed unit interval $[0,1]$ in
	$$
		[0,1]\cup\Bigl(\bigcup_{n\in\N}\{-1/n\}\Bigr)
	$$
	would lie in no ``component''!
	{\bf Components are closed}, however,
	for a component together with any of its limit points which it did
	not contain would be
	connected and properly contain the original.
	Any non-empty subset which is both open and closed is a union of components of
	the topological space.

	It turns out to be useful to use the concept of local finiteness of
	a collection of subsets of the boundary: we say that
	the collection \seq{\cU}\a, $\cU_\a\sub X$, is {\em locally finite\/} if
	each $p\in X$ has an open neighbourhood
	$\cW$ which intersects only finitely many of the $\cU_\a$.
	A moment's thought shows that the concept of
	local finiteness of a collection $\seq{\cU}\a,\,\cU_\a\sub Y\sub X$,
	is independent of whether one takes the open neighbourhoods $\cW$ to
	be sets of the $X$-topology or the relative $Y$-topology.  In the
	following we will choose the former as this fits more with one's
	intuitive idea of a ``neighbourhood'' of a boundary component.

\begin{lemma}
	\thmlabel{isol comps}
	\strut\par
	\begin{enumerate}\thisenumparens{alph}
	\item \enlabel{comp lem 1}
		An isolated boundary set $B$ is a union of components of $\d_\phi$
		(in the latter's relative topology).
	\item \enlabel{comp lem 2}
		The boundary $\d_\phi$ consists of a locally finite
		collection of boundary components if and only if one can find disjoint
		$\Mhat$-open neighbourhoods of each component.  Any
		boundary component which is compact is thereby an isolated
		boundary set.
	\item \enlabel{comp lem 3}
		If $B=\bigcup B_i$ is an isolated boundary set, each $B_i$ is
		a boundary component, and the
		$B_i$ are separated by $\Mhat$-open neighbourhoods $\cU_i$,
		all of which are disjoint, then $B$ is
		a union of a {\em finite\/} number of boundary components.
	\end{enumerate}
\end{lemma}
\begin{remarks}
	Even though an isolated boundary set is necessarily
	compact, it may not be
	expressible as a {\em finite\/} union of components of $\d_\phi$.
	Consider the boundary set $B=\d_\phi$ defined by
	\begin{equation}
		\eqlabel{nasty comps}
		\left\{\begin{array}{l}
			\d_\phi=X\times\{0\},\hbox{ where }
				X=[0,1]\cup\Bigl(\bigcup_{n\in\N}\{-1/n\}\Bigr)\\
			\multicolumn{1}{r}{\Mhat=\R^2,\ \phi(\cM)=\Mhat\sm\d_\phi}.
		\end{array}\right.
	\end{equation}
	Then $B$ is isolated but consists of an infinite number of boundary
	components.

	Furthermore, a union $B$ of compact components of the boundary
	$\d_\phi$ (even if these are finite in number)
	may not be isolated (although
	\enref{comp lem 2}
	shows that local finiteness of the collection of components of $\d_\phi$ is
	a sufficient condition for $B$ to be isolated, if it is compact).
	Every open neighbourhood of the compact component $[0,1]\times\{0\}$
	of $\d_\phi$ in
	\eqref{nasty comps}
	above has non-empty intersection with
	$\d_\phi\sm([0,1]\times\{0\})$, and so
	is not isolated.

	Finally, we note that ``locally finite'' may, of course, be replaced
	by ``finite'' in the statement of \enref{comp lem 2} above.
\end{remarks}
\expandafter\ifx\csname pf\endcsname\relax
	\begin{proof}[of Lemma~\thmref{isol comps}]
\else
	\begin{proof}[Lemma~\thmref{isol comps}]
\fi
	To show
	\enref{comp lem 1},
	note that $B$ is compact, so
	$B$ is closed in $\Mhat$ and thus in $\d_\phi$.  Since $B$ is isolated one
	sees from the definition that it is also open in $\d_\phi$.
	It follows that it must be a union of components of $\d_\phi$.

	We next show the ``only if'' clause of
	\enref{comp lem 2}.
	Let $\seq BA_{A\in\cA}$ represent the locally finite
	collection of boundary components.  Consider any member of this
	collection, $B_A$.
	Since $B_A$ is a boundary component, it is a closed subset of both
	$\d_\phi$ and $\Mhat$.
	Now define
	$$
		X_A \defeq \bigcup_{\stackrel{\scriptstyle C\in\cA}{C\ne A}} B_C.
	$$
	Suppose that $p\in\overbar{X_A}\sm X_A$, which implies that $p\in B_A$.
	Since $p\in\overbar{X_A}$, there is a sequence $\seq pi$ of
	points of $X_A$ such that $p_i\to p$.
	Now because the collection of boundary components is locally finite,
	there is an open neighbourhood $\cW$ of $p$ such that $\cW$ intersects
	only finitely many boundary components: let these (distinct)
	components be $B_A$, $\range B{C_1}{C_m}$.
	It follows that there is a subsequence of the \seq pi lying in one
	of the $B_{C_j}$, whence $p\in\overbar{B_{C_j}}$.
	Since $B_{C_j}$ is closed, $p\in{B_{C_j}}$ and so $p\in X_A$,
	which yields a contradiction.
	Thus $X_A$ is closed, and $X_A$ and $B_A$ are disjoint closed sets.

	Now let $d$ be a Riemannian metric on $\Mhat$.  For $p\in B_A$,
	define
	$$
		\mu_{A,p} \defeq \inf \set d(p,x) for x\in X_A \eset.
	$$
	Clearly $\mu_{A,p}\ge0$; assume that $\mu_{A,p}=0$.
	This means that there is an infinite sequence  \seq xi
	of points in $X_A$ such that $d(p,x_i)\to0$ as $i\to\infty$.
	That is, $x_i\to p$ as $i\to\infty$ which implies that
	$p\in\overbar{X_A}=X_A$.  This is a contradiction since $B_A$ and
	$X_A$ are disjoint, and thus $\mu_{A,p}>0$.

	Now define $\e_{A,p}\defeq\mu_{A,p}/10$,
	$\cV_{A,p}\defeq\set q\in\Mhat for d(p,q)<\e_{A,p}\eset$ and
	$$
		\cU_A\defeq\bigcup_{p\in B_A}\cV_{A,p}.
	$$
	Clearly, $\cU_A$ is an open neighbourhood of $B_A$.

	Consider the collection of open neighbourhoods $\seq{\cU}A_{A\in\cA}$.
	Suppose that $r\in\cU_C\cap\cU_D$ ($C\ne
	D$).  This implies that $r\in\cV_{C,p}$ for some $p\in B_C$
	and $r\in\cV_{D,q}$ for some $q\in B_D$.
	Without loss of generality assume that $\e_{C,p}\ge\e_{D,q}$, and then
	\begin{eqnarray*}
		d(p,q)
		\ale d(p,r) + d(q,r)\\
		\alt \e_{C,p}+\e_{D,q}\\
		\ale 2\e_{C,p}\\
		\eq\mu_{C,p}/5.
	\end{eqnarray*}
	This contradicts that $\mu_{C,p}$ is the greatest lower bound
	of $\set d(p,x) for x\in X_C\eset$.
	Thus $\seq{\cU}A_{A\in{\cA}}$ is a collection of disjoint
	$\Mhat$-open neighbourhoods of the boundary components
	$\seq{B}A_{A\in{\cA}}$.
	Clearly any boundary component $B_A$ which is compact must be an isolated
	boundary set since $\cU_A\cap\d_\phi=B_A$.

	To show the ``if'' part of
	\enref{comp lem 2}:
	let $p\in\d_\phi$.  Now $p$ lies in a particular boundary component
	$B_A$.  Thus $\cU_A$ is an open neighbourhood of $p$ in $\Mhat$ and
	$\cU_A\cap\d_\phi$ is open in the relative topology of $\d_\phi$.
	Also
	$\cU_A\cap\d_\phi=B_A$,
	demonstrating that
	$\seq BA_{A\in\cA}$
	is indeed a locally finite collection of boundary components.

	Finally, \enref{comp lem 3} is immediate.
	The disjoint $\Mhat$-open neighbourhoods $\cU_i$
	of the boundary components
	$B_i$ form an open cover $\seq{\cU}i$ of $B$.
	Since $B$ is an isolated boundary set, it is compact.  The open
	cover $\seq{\cU}i$
	must therefore admit a finite subcover, implying that $B$ is a union
	of a {\em finite\/} number of boundary components.
\end{proof}

\subsection{Invariance}

The utility of the concept of an isolated boundary set is seen through
the following theorem, which establishes that the property of being
isolated is invariant under boundary set equivalence.

\begin{theorem}
	\thmlabel{isol invar}
	If $B\sub\d_\phi$ is an isolated boundary set, then any $B'\sim B$ is
	also an isolated boundary set.
\end{theorem}
\begin{proof}
	$B'$ is compact by Theorem~\thmref{compactness stable}.  Assume that
	it is not isolated.

	Since $B$ is isolated,
	there is an open neighbourhood (in $\Mhat$) $\cU$ of $B$ such that
	$\cU\cap\d_\phi= B$.
	Now let $d$ be a complete Riemannian metric on $\Mhat$.
	For each $b\in B$, there is a real $\a_b>0$ such that $\tilde B_b\sub \cU$,
	where
	$$
		\tilde B_b\defeq\set x\in\Mhat for d(b,x)<\a_b\eset.
	$$
	Let
	$$
		B_b\defeq\set x\in\Mhat for d(b,x)<\a_b/2\eset.
	$$
	So $\cU*=\bigcup_{b\in B} B_b$ is an open neighbourhood of $B$ in
	$\Mhat$.

	Now since $B$ is compact, there is a finite subcover
	$\{B_{b_1},\ldots,B_{b_m}\}$
	of the collection
	$\set B_b for b\in B\eset$.
	Of course $A\defeq\bigcup_{i=1}^m B_{b_i}$ is an open neighbourhood of $B$ in
	$\Mhat$. Further, $\bar A=\bigcup_{i=1}^m\bar B_{b_i}$ is compact.

	Let $x\in\bar A\sm B$: say $x\in \bar B_{b_i}$.
	Then $x\in \tilde B_{b_i}$, so $x\in \cU$.  Since $x\nin B$ and
	$\cU\cap\d_\phi= B$, $x\nin\d_\phi$, and it follows that $\bar
	A\cap\d_\phi=B$.

	Since $B\rhd B'$, there is an open neighbourhood $\cV'$ of $B'$ in
	$\Mhat'$ such that
	$$
		\phi\circ(\phi'\inv)(\cV'\cap\phi'(\cM)) \sub A.
	$$
	Now $B'$ is not isolated so there is a boundary point
	$\circover b'\in\d_{\phi'}\sm B'$ which lies in $\cV'$.  Let \seq pi be a
	sequence of points
	in $\cM$ with $\phi'(p_i)\to \circover b'$. There exists an $N\in\N$
	such that $\phi'(p_i)\in \cV'$ for all $i\ge N$, whence $\phi(p_i)\in
	A\sub\bar A$ ($\forall\; i\ge N$).
	Since $\bar A$ is compact, the sequence $\{\phi(p_i)\}$ must have a
	limit point $y\in\bar A$.
	One can show in a straightforward manner that $y\nin\phi(\cM)$ and
	that $y\nin\Mhat\sm \overbar{\phi(\cM)}$.  That is, $y\in\d_\phi$, so $y\in
	B$.
	Now since $B'\rhd B$, the sequence $\{\phi'(p_i)\}$ must have a
	limit point in $B'$
	\cite[Theorem~19]{Scott94}, but $\phi'(p_i)\to \circover b'\nin B'$, a
	contradiction.

	Thus $B'$ is an isolated boundary set.
\end{proof}

\section{Connected boundary sets}
\seclabel{connectedness}

Unlike compactness and isolation, connectedness of boundary sets is not
invariant under boundary set equivalence.  Examples of this are easy to
construct, in any dimension. We proceed to do so next even for {\em isolated\/}
boundary sets.

\begin{example}\backtoroman
	\label{ex:arb genus ex}
	This example shows that for any dimension $n\ge2$ one can find
	connected, isolated boundary sets equivalent to boundary sets with
	{\bf\em any\/} number $g$ of connected components
	(where $g$ is a positive integer).

	First consider the $n=2$ case.
	Let $\Mhat$ be a compact, connected surface of genus $g$, where $g$
	is an integer greater than 0 (so $\Mhat$ is the
	surface of a sphere
	with $g$ ``handles'', or rather $g$ ``holes''---see
	Figure~\figref{arb genus ex} where the case $g=4$ is depicted; the
	holes should be arranged in a ring as shown in the Figure).
	The case $g=2$ is depicted in Figure~\figref{arb genus ex (b)}.
	\begin{figure}[p]
		\kern.3cm
		\centering\leavevmode\epsfbox{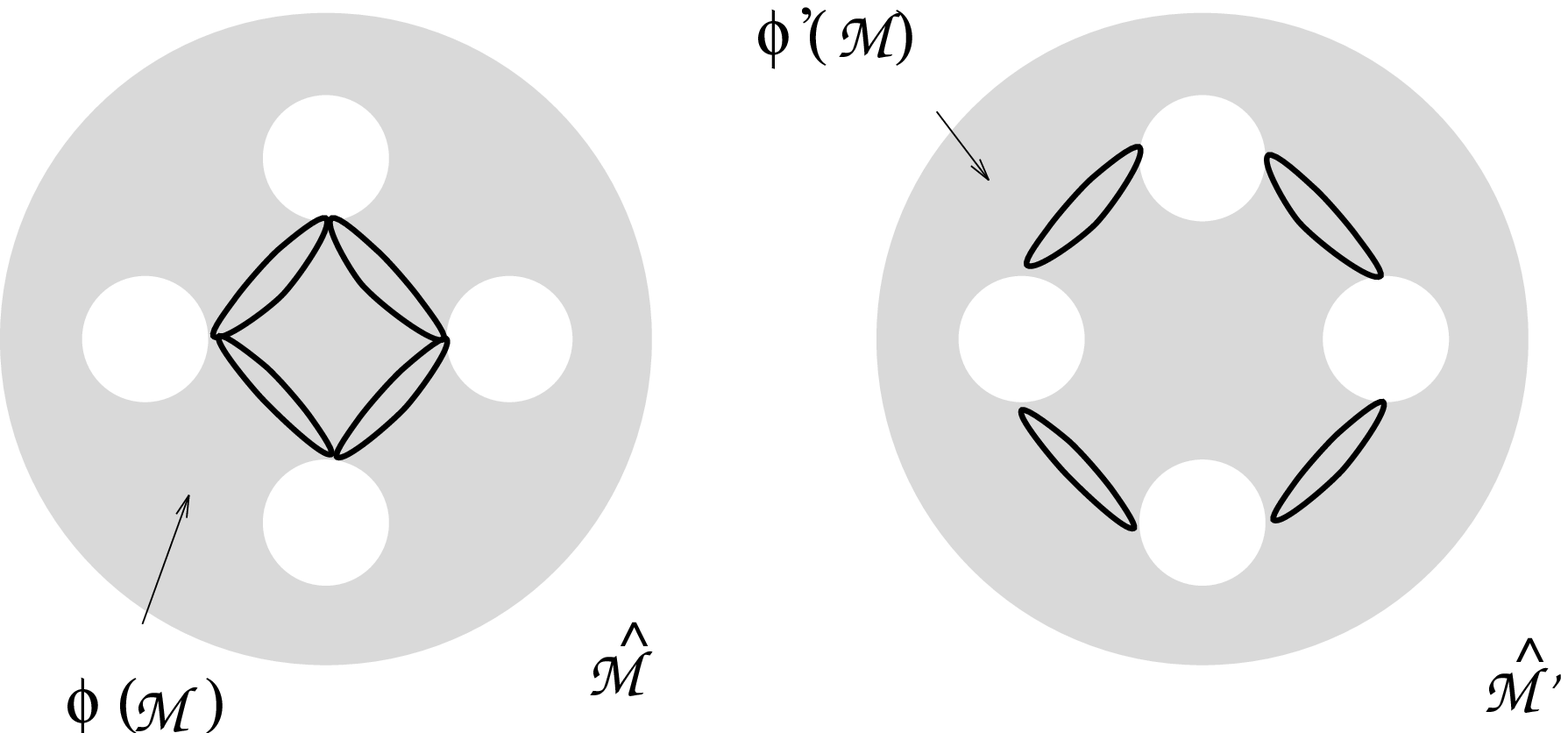}
		\caption{A schematic representation of the
			compact, connected surface of genus $4$ as discussed in
			Example~\protect\ref{ex:arb genus ex}.
			The shaded areas represent one side of the surface, and the
			black lines represent the boundary sets in question.
			\figlabel{arb genus ex}}
	\end{figure}
	\begin{figure}[p]
		\centering\leavevmode\epsfbox{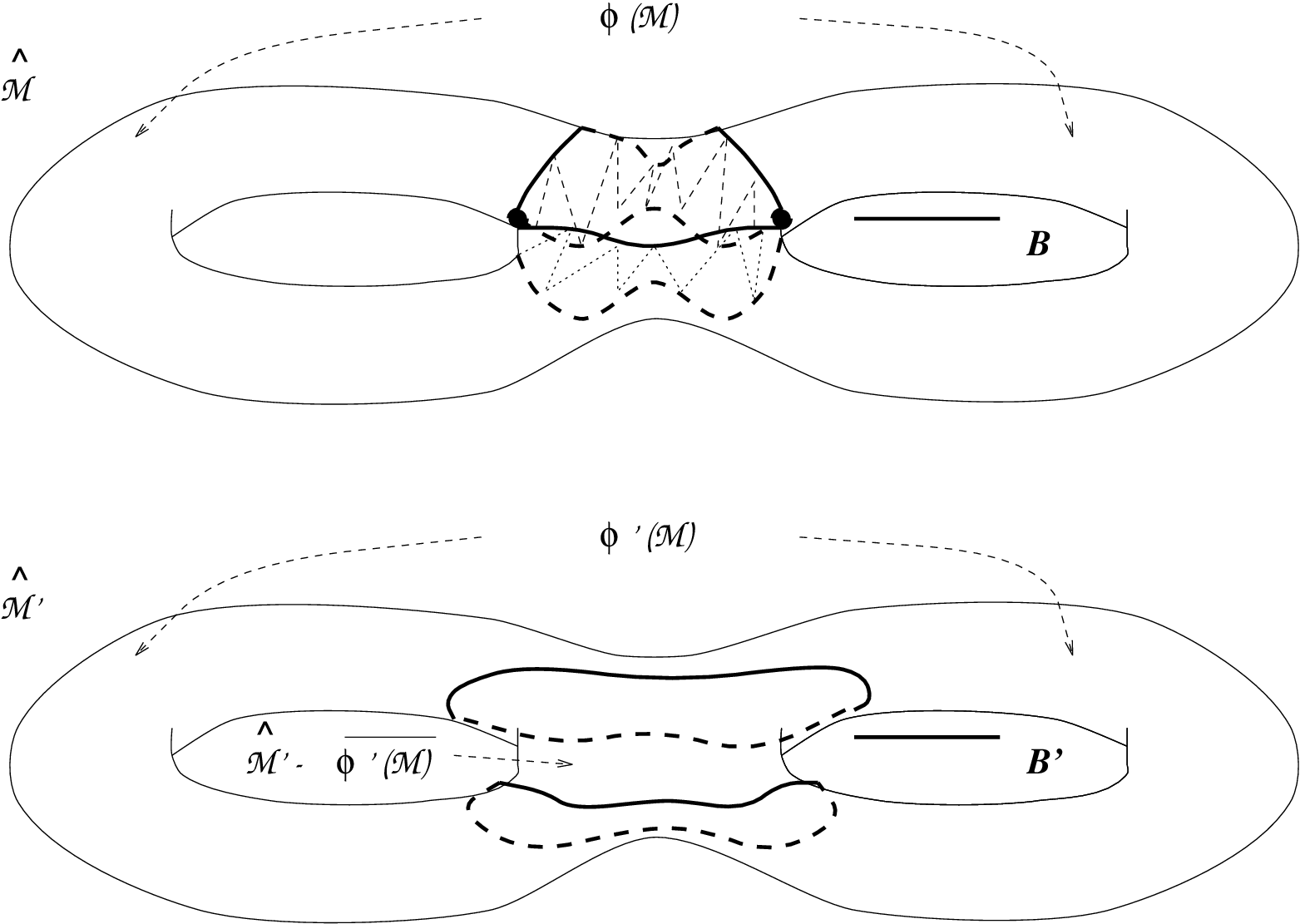}
		\caption{The $g=2$ case of
			Example~\protect\ref{ex:arb genus ex}.
			\figlabel{arb genus ex (b)}}
	\end{figure}

	Let $B$ consist of $g$ embedded circles
	(see the Figures). If $g>2$,
	each circle is to intersect precisely two neighbours, at one point
	each: each circle ``clothes an arm'' of the multi-torus.
	If $g=2$, there is only one neighbouring circle,
	but there are {\em two\/} intersection points. In the degenerate
	$g=1$ case there are, of course, no neighbours.

	Let $\cM$ be the component of $\Mhat\sm B$
	which does not
	include those portions of the surface in the centre of $\Mhat$---see
	the Figures.
	Let $\phi$ be the inclusion. Then $B=\d_\phi$ is a
	connected, compact and indeed isolated boundary set.

	Let $\Mhat'=\Mhat$ and $\phi'$ be a re-embedding
	of $\phi(\cM)$ onto $\phi'(\cM)$, where each ``sleeve'' of
	$\phi(\cM)$, terminating in a circle of $B$,
	retracts away from the centre of
	$\Mhat'$.
	Then $B'=\d_{\phi'}$ has $g$ components, and clearly we
	have $B\sim B'$.

	For $n>2$, consider (with the same notation as above) the
	envelopments
	$(\cM\times X_0,\Mhat\times X_0,\psi\defeq\phi\times{}$identity$)$
	and
	$(\cM\times X_0,\Mhat'\times X_0,\psi'\defeq\phi'\times{}$identity$)$,
	where $X_0$ is any connected, compact manifold.  Since
	$\d_\psi=\d(\phi(\cM)\times X_0)$ is just
	$\d_\phi\times X_0$ (as $X_0$ has no boundary, of course),
	and  a product of connected, compact sets is connected and
	compact,
	$\d_\psi$ is a connected, isolated boundary set.  Similarly,
	$\d_{\psi'}$ is an isolated boundary set, but consists of $g$
	connected components.  Now since $\d_\psi\sim\d_{\psi'}$, we have
	the desired example of a connected, isolated boundary set
	equivalent to an isolated boundary set with any number of connected
	components, for any dimension $n\ge2$.
\end{example}

\subsection{The connected neighbourhood property}

\begin{sloppypar}
	Example~\ref{ex:arb genus ex} demonstrates that
	connectedness does not ``pass to'' arbitrary equivalence classes of
	boundary sets.
	If, however, a boundary set $B$ of an envelopment $(\cM,\Mhat,\phi)$
	satisfies a certain very easily visualised property (the {\em connected
	neighbourhood property\/}), then one can show that
	{\em all\/} equivalent boundary sets satisfy this same property.
	Furthermore, all these equivalent
	boundary sets must themselves be connected
	(see Theorem~\thmref{CNP is connected}).
\end{sloppypar}

It so happens that this property can very
easily be generalised and yet remain invariant under boundary set
equivalence (see
Theorem~\thmref{TNP invar}).
Below, we define the
``connected neighbourhood property'' of boundary sets to be a special
case of the ``$T$ neighbourhood property'' of boundary sets,
where $T$ is a given topological property.

\begin{definition}\backtoroman
	A {\em topological property\/} $T=T(X)$
	of a topological space $X$ is a property of $X$ such that
	all topological spaces $Y$ which are homeomorphic to $X$ have the
	same property: i.e.,
	$T(X)\implies T(Y)$ for all $Y$ homeomorphic to $X$.%
	\footnote{
		Actually we could let $T$ be a
		property of {\em differentiable manifolds\/} which is invariant under
		{\em diffeomorphisms}.  In particular, properties of the
		tangent bundle of the manifold could then be used here.
	}
\end{definition}

\begin{definition}\backtoroman
	Let $B$ be a boundary set of an envelopment $(\cM,\Mhat,\phi)$.
	An open neighbourhood $\cU$ (in $\Mhat$) of $B$
	shall be called
	{\em $T$-nice\/} if
	its intersection with $\phi(\cM)$ satisfies $T$ (in the relative topology of
	$\cU\cap\phi(\cM)$); i.e., if $T(\cU\cap\phi(\cM))$ is true.
\end{definition}

\begin{definition}\backtoroman
	Using the notation of the previous definition,
	we will say that $B$ satisfies the
	{\em $T$ neighbourhood property\/} if
	all open neighbourhoods $\cU$ of $B$ in $\Mhat$ contain
	a $T$-nice, open neighbourhood
	$\cV$ of $B$ (in $\Mhat$).
	In the following, for the sake of brevity, we will
	abbreviate ``$T$ neighbourhood property'' to ``\TNP''.
\end{definition}

In this section we will concentrate on the case
$$
	T(X)={}\hbox{``$X$ is connected''}{}\defeq C(X).
$$
Thus
we have defined the ``connected neighbourhood property'', which, in
keeping with the abbreviation ``\TNP'', we
shall denote by ``\CNP''.

We now prove that a
boundary set $B$ satisfying the \CNP{} is always connected.

\begin{theorem}
	\thmlabel{CNP is connected}
	Let $B$ be a
	boundary set of an envelopment $(\cM,\Mhat,\phi)$.
	If $B$ satisfies the \CNP, then $B$ is connected.
\end{theorem}
\begin{proof}
	Suppose that $B$ is disconnected.  Then there exist non-empty,
	open sets $\cU$ and $\cV$ of $\Mhat$ such that
	\begin{itemize}
	\item
		$\cU\cup\cV$ is an open neighbourhood of $B$,
	\item
		$\cU\cap B\ne\es$ and $\cV\cap B\ne\es$, and
	\item
		$X\defeq \cU\cap B$ and $Y\defeq \cV\cap B$ are disjoint.
	\end{itemize}
	That is, $X$ and $Y$ are disjoint, non-empty
	open sets of the relative topology
	on $B$, with $X\cup Y=B$.

	Let $d$ be a Riemannian metric on $\Mhat$.
	For any $x\in X$, $y\in Y$, define
	$$
		\p(x)\defeq\inf\set d(x,y) for y\in Y\eset,
		\qquad
		\p(y)\defeq\inf\set d(y,x) for x\in X\eset.
	$$
	Clearly these functions take values in the non-negative reals.
	We claim that they cannot vanish at any point.  Assume that this is
	not true: without loss of generality,
	say that $\p(x)=0$ for some $x\in X$.
	This implies that there is a sequence $\seq ym$ of points in $Y$ for which
	$d(x,y_m) \mathop{\stackrel{m\to\infty}\longrightarrow} 0$.
	Since $\cU$ is open, there is some $\mu_x>0$ such that
	$B_d(x,\mu_x)\sub \cU$.
	This means that
	$B_d(x,\mu_x)\cap B\sub X$. (Similarly, we define $\mu_y$ for $y\in Y$ to
	be some positive number for which $B_d(y,\mu_y)\sub \cV$, so that
	$B_d(y,\mu_y)\cap B\sub Y$).  It follows, then, that $y_m\in X$ eventually,
	contradicting the disjointness of $X$ and $Y$.  Thus $\p(x),\p(y)>0$,
	$\forall\:x\in X,y\in Y$.

	For any $x\in X$, $y\in Y$, define
	$$
		\e_x\defeq\min\{\p(x)/10,\mu_x\},
		\qquad
		\e_y\defeq\min\{\p(y)/10,\mu_y\}.
	$$
	Clearly $B_d(x,\e_x)\sub \cU$ and $B_d(y,\e_y)\sub \cV$.
	Let
	$$
		\tilde \cU = \bigcup_{x\in X} B_d(x,\e_x)
		\qquad\hbox{and}\qquad
		\tilde \cV = \bigcup_{y\in Y} B_d(y,\e_y).
	$$
	Assume that $p\in\tilde \cU\cap\tilde \cV$, so
	$p\in B_d(x,\e_x)$ for some $x\in X$ and
	$p\in B_d(y,\e_y)$ for some $y\in Y$.  Now
	\begin{eqnarray*}
		d(x,y) \ale d(x,p) + d(p, y)\\
		\alt 2\max\{\e_x,\e_y\}.
	\end{eqnarray*}
	Without loss of generality, let us say that $\e_x\ge\e_y$.  Then
	$d(x,y)< 2\e_x\le\p(x)/5$.  This is a contradiction, so that
	$\tilde \cU$ and $\tilde \cV$ must be disjoint.

	Now $\tilde \cU\cup\tilde \cV$ is an open neighbourhood of $B$ in $\Mhat$,
	and $\tilde \cU\cap B$ and $\tilde \cV\cap B$ are non-empty and disjoint.
	Since $B$ satisfies the \CNP,
	there exists an open neighbourhood $\cW$ of $B$ with $\cW\sub\tilde
	\cU\cup\tilde \cV$, for which $\cW\cap\phi(\cM)$ is connected.  Now
	$\cW\cap\tilde \cV\ne\es$, for were the intersection empty,
	we would have $\cW\sub \tilde \cU$, which, since $\cW$
	is an open neighbourhood
	of $B$, would contradict the fact that $\tilde \cV\cap B\ne\es$.
	Similarly, $\cW\cap\tilde \cU\ne\es$.  It follows that
	$\cW\cap\tilde \cU\cap\phi(\cM)$ and $\cW\cap\tilde \cV\cap\phi(\cM)$
	are two disjoint, non-empty open sets, which contradicts that
	$\cW\cap\phi(\cM)$ is connected.

	\nobreak
	Thus $B$ must be connected, and the result is proven.
\end{proof}

\subsection{The $T$ neighbourhood property and invariance}

We now proceed to a result with very wide applicability.  It says that
for {\bf\em any\/} topological property $T$, if a boundary set $B$
satisfies the
$T$ neighbourhood property (\TNP), then {\em all\/} boundary sets which are
equivalent to $B$ must also satisfy this property.

To begin with, we need the following lemma, which gives an equivalent
formulation of the covering relation
to that given in
\S\secref{intro} (Definition~14 of~\citeasnoun{Scott94}).

\begin{lemma}
	\thmlabel{equiv covering}
	$B\rhd B'$ if and only if for any open neighbourhood $\cU$ of $B$ in
	$\Mhat$, there is an open neighbourhood $\cU'$ of $B'$ in
	$\Mhat'$ with
	$$
		\phi\circ(\phi')^{-1} (\cU'\cap\phi'(\cM)) = \cU\cap\phi(\cM).
	$$
\end{lemma}
\begin{proof}
	Sufficiency is immediate, because
	for any open neighbourhood $\cU$ of $B$ in
	$\Mhat$, there is an open neighbourhood $\cU'$ of $B'$ in
	$\Mhat'$ with
	\begin{eqnarray*}
		\phi\circ(\phi')^{-1} (\cU'\cap\phi'(\cM)) \eq \cU\cap\phi(\cM)\\
		&\sub&\cU.
	\end{eqnarray*}
	Thus $B\rhd B'$.

	\begin{figure}
		\maxpspicwid=.8\hsize
		\kern.3cm
		\centering\leavevmode\epsfbox{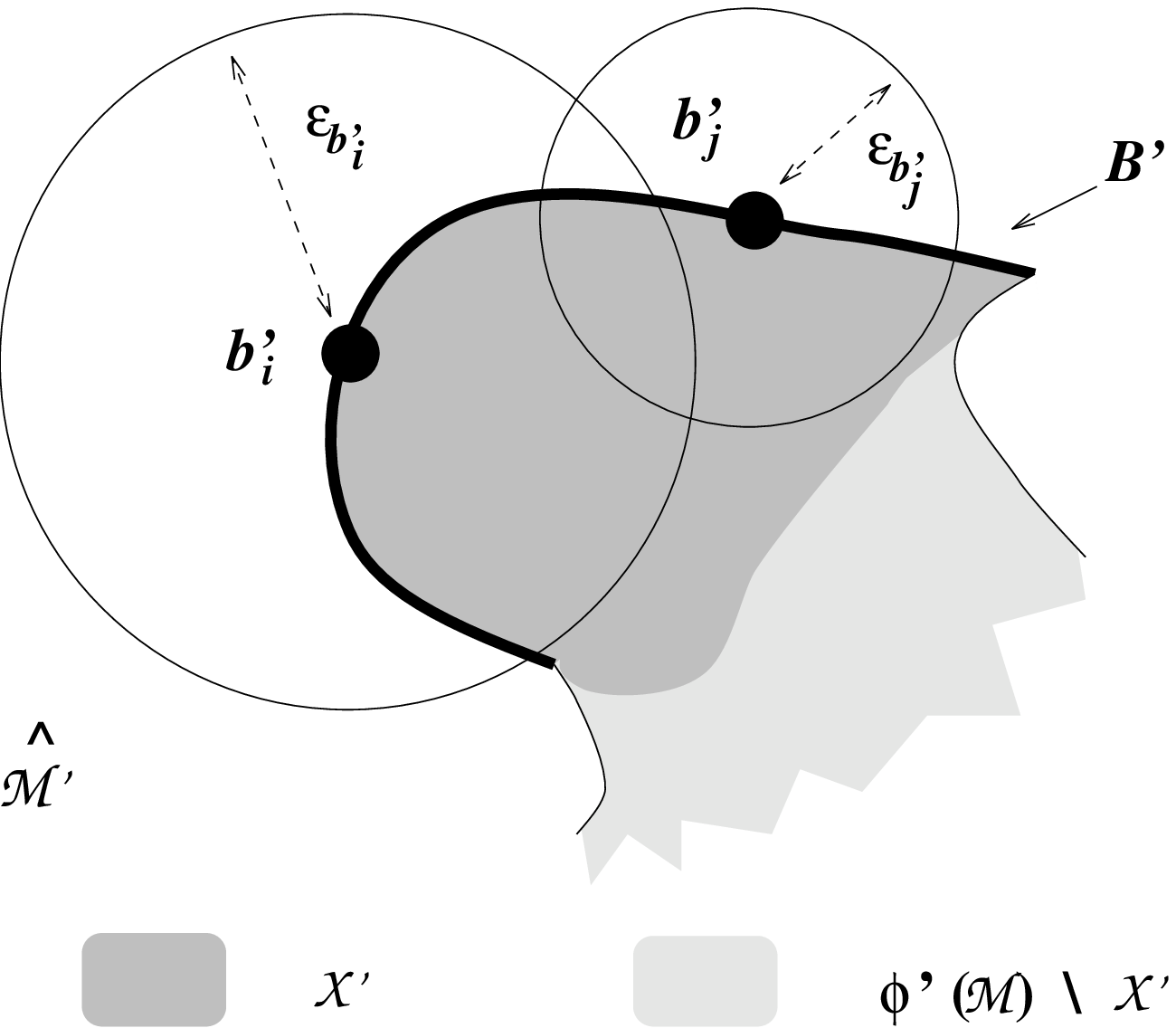}
		\caption{Illustrating the construction of $\cU'$ in the
			proof of
			Lemma~\protect\thmref{equiv covering}.
			\figlabel{equiv covering}}
	\end{figure}
	To show necessity, let $B\rhd B'$; i.e.,
	for any open neighbourhood $\cU$ of $B$ in
	$\Mhat$, there is an open neighbourhood $\cV'$ of $B'$ in
	$\Mhat'$ with
	$$
		\phi\circ(\phi')^{-1} (\cV'\cap\phi'(\cM)) \sub \cU.
	$$
	Let $d'$ be a Riemannian metric on $\Mhat'$ and let
	$$
		X'\defeq\phi'\circ\phi^{-1} (\cU\cap\phi(\cM)).
	$$
	Now suppose that for some $b'\in B'$, there does not exist an
	$\e_{b'}>0$ such that $B_{d'}(b',\e_{b'})\cap\phi'(\cM)\sub X'$.
	Thus a sequence \seq pn of points of $\cM$ can be found such that
	$\phi'(p_n)\nin X'$, but
	$\phi'(p_n)\mathop{\stackrel{n\to\infty}\longrightarrow}b'$.
	Since $B\rhd B'$, by Theorem~19 of \citeasnoun{Scott94},
	\seq pn admits a subsequence $\{p_{n_i}\}$ such that
	$\phi(p_{n_i})\mathop{\stackrel{i\to\infty}\longrightarrow}b\in B$
	for some $b$. So $\phi(p_{n_i})$ must eventually lie in $\cU$
	which contradicts that $\phi'(p_{n_i})\nin X'$.

	Thus for each $b'\in B'$, there exists an $\e_{b'}>0$ such that
	$B_{d'}(b',\e_{b'})\cap\phi'(\cM)\sub X'$.
	(See Figure~\figref{equiv covering}.)
	Define
	$$
		\cU' \defeq X' \cup \biggl(\bigcup_{b'\in B'}
			B_{d'}(b',\e_{b'})\biggr).
	$$
	Then $\cU'$ is an open neighbourhood of $B'$ satisfying the required
	property that
	$$
		\phi\circ(\phi')^{-1} (\cU'\cap\phi'(\cM)) = \cU\cap\phi(\cM).
	$$
\end{proof}

Using this equivalent formulation of ``covering'', the main theorem can
now be proven in a straightforward manner.

\begin{theorem}
	\thmlabel{TNP invar}
	Let $T$ be a topological
	property.
	If $B$ satisfies the \TNP{} and $B\sim B'$,
	then $B'$ also satisfies the \TNP{}.
\end{theorem}
\begin{proof}
	Assume that $B$ satisfies the \TNP{} and $B\sim B'$,
	but that $B'$ does not satisfy the \TNP.
	This means that there is an open neighbourhood $\cU'$ of $B'$ in
	$\Mhat'$ such that $\cU'$ contains no $T$-nice, open neighbourhoods of
	$B'$. Since $B\lhd B'$, there is
	an open neighbourhood $\cU$ of $B$ in
	$\Mhat$ such that
	\begin{equation}
		\eqlabel{TNP one way}
		\phi'\circ\phi^{-1} (\cU\cap\phi(\cM)) \sub \cU'.
	\end{equation}
	Since $B$ satisfies the \TNP, there is a $T$-nice, open neighbourhood
	$\cV$ of $B$ contained in $\cU$.  By
	Lemma~\thmref{equiv covering},
	since
	$B\rhd B'$, there is
	an open neighbourhood $\cV'$ of $B'$ with
	\begin{equation}
		\eqlabel{TNP other way}
		\phi\circ(\phi')^{-1} (\cV'\cap\phi'(\cM)) = \cV\cap\phi(\cM).
	\end{equation}
	Now define $\cW'\defeq(\cV'\cap\cU')\cup(\cV'\cap\phi'(\cM))$.
	It is clear that $\cW'$ is an open neighbourhood of $B'$,
	$\cW'\sub\cU'$, and $\cW'\cap\phi'(\cM)= \cV'\cap\phi'(\cM)$.

	Thus we have constructed an open neighbourhood $\cW'$ of $B'$ which is
	contained in $\cU'$, and satisfies
	\eqref{TNP other way}
	with $\cV'$ replaced by $\cW'$.
	Since $\cV$ is $T$-nice and
	$\phi'\circ\phi^{-1}$ is a homeomorphism, it follows that
	$\cW'$ is $T$-nice.%
	\footnote{
		In fact, $\phi'\circ\phi^{-1}$ is a smooth diffeomorphism.
		Thus, as noted in an earlier footnote, were $T$ a
		property of differentiable manifolds invariant under
		diffeomorphisms, exactly the same argument would apply.
	}
	This is a contradiction, and the theorem is
	proven.
\end{proof}

\begin{corollary}
	\thmlabel{TNP well-def}
	If\/ $T$ is a topological property, then the \TNP{}
	is a well-defined property
	of abstract boundary sets $[B]$: we say that $[B]$
	satisfies the \TNP{} if and only if  the representative boundary set
	$B$ satisfies the \TNP.
\end{corollary}
\begin{proof}
	By
	Theorem~\thmref{TNP invar},
	this definition is independent of the representative boundary set
	chosen.
\end{proof}

\begin{remark}
In particular, the \TNP{} passes to the abstract boundary $\cB(\cM)$.
\end{remark}

\smallskip
For the moment we let $T={}$``connected'', and investigate the \CNP.
First we find examples of connected boundary sets $B$ not satisfying the \CNP.

\begin{examples}\backtoroman
\item
	\label{conn, not CNP}
	Consider the isolated boundary set $B$ of
	Example~\ref{ex:arb genus ex}.
	There, $B$ is connected, but is readily seen by inspection not to
	satisfy the \CNP.

	Alternatively, $B\sim B'$, where $B'$ consists of $g$ components and
	thus cannot satisfy the \CNP{} by
	Theorem~\thmref{CNP is connected}.
	It then follows from
	Theorem~\thmref{TNP invar}
	that $B$ cannot satisfy the \CNP{} either.

\item
	\label{conn, non-comp, SC, not CNP}
	In this example, the boundary set in question is non-compact,
	and thus not isolated.
	Let $\Mhat=\R \times S^1=\set (t,\th) for t\in\R, -\pi<\th\le\pi\eset$.
	Define $\cM\defeq\Mhat\sm\{\th=\pi\}$,
	and let $\phi$ be the inclusion.
	Then the boundary set $B=\d_\phi=\{\th=\pi\}$ is connected.

	Let $\cU$ be an open neighbourhood of $B$, given by
	$$
		\cU \defeq \set (t,\th) for t\in\R, -\pi<\th<-a \hbox{ or }
		a<\th\le\pi\eset,
	$$
	where $0<a<\pi$.  Then $\cU\cap\phi(\cM)$ has precisely two
	components.  Furthermore, for any open neighbourhood $\cV$ of $B$
	contained in $\cU$,
	$\cV\cap\phi(\cM)$ must have at least two components.  Thus $B$ does
	not satisfy the \CNP.

	We note that a similar argument to that used at the end of
	Example~\ref{conn, not CNP}
	could be used here.
	If $\Mhat'=\Mhat$ and $\phi'$ maps $(t,\th)$ to
	$(t',\th')=(t,\th/2)$, then
	$$
		B'=\d_{\phi'}=\{\th'=-\pi/2\}\cup\{\th'=\pi/2\}
	$$
	is the union of two components, and $B'\sim B$.
\end{examples}

\subsection{Isolated boundary points and the \protect\CNP}

We now show that, for $n\ge2$,
all boundary sets equivalent to an {\em isolated\/}
boundary point must satisfy the \CNP.  There are, of course, many other
types of boundary point which satisfy the \CNP---for specific examples
this question is often easily settled by inspection.

In the following one wishes to talk about balls of certain radii about
an isolated boundary point, so again we fix a Riemannian metric $d$ for
$\Mhat$.

\begin{lemma}\thmlabel{isol in sense that...}
	Let $n\ge2$. If $p$ is an isolated boundary point then, for some
	$\e>0$, there is an open metric ball $B_\e$ about $p$, of radius
	$\e$,  which
	intersects $\d_{\phi}$ at \rom(and only at\rom) $p$,
	and is  such that $B_\e\sm\{p\}\sub\phi(\cM)$.
	It follows that, if $n\ge2$, an isolated boundary point satisfies
	the \CNP.
\end{lemma}
\begin{proof}
	Since $p$ is an isolated boundary point, there is an open neighbourhood
	$\cU$ of $p$ in $\Mhat$ such that $\cU\cap\d_\phi=\{p\}$.
	Let $B_\e\defeq B_d(p,\e)$ be an open $d$-ball about $p$  with radius
	$\e>0$, such that $B_\e\sub\cU$.  Clearly $B_\e\cap\d_\phi=\{p\}$.

	For $n\ge2$, $B_\e\sm\{p\}$ is connected.  We already know that
	$(B_\e\sm\{p\})\cap\d_\phi=\es$,
	and $X\defeq(B_\e\sm\{p\})\cap\phi(\cM)\ne\es$,
	since $p$ is a boundary point and so must be a limit point of a
	sequence of points in $\phi(\cM)$.
	Also, $X$ is open in the relative topology of $B_\e\sm\{p\}$,
	since $\phi(\cM)$ is open.

	Suppose that
	$Y\defeq(B_\e\sm\{p\})\cap(\Mhat\sm\overbar{\phi(\cM)})\ne\es$.
	It is readily seen that $Y$ is also open in the relative topology
	of $B_\e\sm\{p\}$, since $\Mhat\sm\overbar{\phi(\cM)}$
	is open.  This yields a contradiction, since $B_\e\sm\{p\}=X\cup Y$
	is the disjoint union of two non-empty open sets, yet
	it is connected. It follows that $B_\e\sm\{p\}\sub\phi(\cM)$.

	For all open $d$-balls $B_\mu\defeq B_d(p,\mu)$, $0<\mu<\e$,
	it is clear that $B_\mu\sm\{p\}$ is a subset of $\phi(\cM)$
	and is connected. Since any open neighbourhood of $p$ in $\Mhat$
	contains such a ball, it follows that $\{p\}$ satisfies the \CNP.
\end{proof}

\begin{corollary}
	\thmlabel{weak:pt conn}
	If $n\ge2$,
	an isolated abstract boundary point $[p]$ satisfies the \CNP.  Every
	boundary set $B\in[p]$ is thus connected.
\end{corollary}
\begin{proof}
	This follows immediately from
	Lemma~\thmref{isol in sense that...},
	Corollary~\thmref{TNP well-def},
	and
	Theorem~\thmref{CNP is connected}.
\end{proof}

\begin{example}\backtoroman
	Lemma~\thmref{isol in sense that...}
	fails trivially for $n=1$, since if $\cM=\R$,
	$\Mhat=S^1=\set\th for -\pi<\th\le\pi\eset$ and $\phi(t)=2\arctan t$,
	then the boundary point $p$ ($\{p\}=\d_\phi$)
	given by $\th=\pi$ is isolated, yet clearly does
	not satisfy the \CNP.
\end{example}

We close this section with an answer to a question posed in
\citeasnoun{Scott94}.

\begin{example}\backtoroman
	\label{ex:disconn comp isol abs set not cont pt}
	We now can find immediately  an example of an
	equivalence class of compact boundary sets which is {\em not\/}
	an abstract boundary point,
	answering a question of \citeasnoun{Scott94} in the negative.
	Simply take the equivalence class containing an isolated
	boundary set with more than one component
	(examples of such are very easy to come
	by: e.g., excise two disjoint closed discs from a plane
	$\Mhat=\R^2$ to obtain $\cM$).
	For $n\ge 2$, such a set cannot be equivalent to a point:
	by
	Theorem~\thmref{isol invar}
	this point would be isolated, whence
	all boundary sets equivalent to it would be connected
	(Corollary~\thmref{weak:pt conn}).
\end{example}

\section{Simple connectedness and vanishing of higher homotopy groups}
\seclabel{homotopy}

Theorem~\thmref{TNP invar}
can be used for many different topological properties $T$, in addition to
$T=C$ (which was examined in detail in \S\secref{connectedness}).  Consider the
following few examples.
\begin{itemize}
\item
	$T(X)={}$``$X$ is simply connected''${}\defeq SC(X)$.%
	\footnote{
		Recall that a simply connected topological space $X$ is a connected
		space for which all loops (i.e., continuous maps $\l:S^1\to X$) are
		homotopic to a constant map: this means that for all $\l$ there is a
		continuous map $F:S^1\times[0,1]\to X$ with $F(\spdot,0)=\l(\spdot)$
		and $F(\spdot,1)$ equal to a fixed point.
		The collection of homotopy classes of loops can be endowed with a
		natural group structure, and this group is called
		the {\em fundamental group\/} $\pi_1(X)$.
	}
\item
	$T(X)={}$``$H_k(X)$ is the trivial group''${}\defeq HOM_k(X)$,
	$k\in\Z,0\le k\le n$.%
	\footnote{
		$H_k(X)$ denotes the $k$th homology group.
	}
\item
	Since we are interested in neighbourhood properties of subsets $B$
	of a manifold, then the intersection $X$ of an open neighbourhood of $B$
	with $\phi(\cM)$ is itself a manifold, although not necessarily a
	connected one.  Thus we may let
	$T(X)={}$``$X$ is parallelisable''. Recall that a differentiable
	manifold $X$ is called {\em parallelisable\/} if the frame bundle $LX$ of
	$X$ admits a nowhere-vanishing (continuous) section.%
	\footnote{%
		Strictly, this is a {\em diffeomorphism-invariant\/} property of
		{\em differentiable manifolds}, but as previously noted, we can allow
		$T$ to be such a property.
	}
	If $X$ has a pseudo-Riemannian metric, this is the condition that $X$
	admit a globally defined, nowhere-vanishing, continuous
	orthonormal field.
\item
	$T(X)={}$``$X$ is $k$-connected''${}\defeq \Pi_k(X)$,
	$k\in\Z,k\ge1$.%
	\footnote{
		The group $\pi_k(X)$ is the $k$th homotopy
		group of the connected
		topological space $X$, and is in
		one-one correspondence with the collection of homotopy
		classes of continuous functions $S^k\to X$.
		If $\pi_k(X)$ is the trivial group (i.e., if all continuous
		functions $S^k\to X$ are homotopic), then
		$X$ is said to be {\em$k$-connected}.
	}
\end{itemize}
(Note that $\Pi_1(X)$ is, of course, the same as $SC(X)$, and
$HOM_0(X)$ is the same as $C(X)$.)
Thus we have defined the \SCNP, \PIkNP{} and \HOMkNP{} of boundary
sets.
Corollary~\thmref{TNP well-def},
of course, shows that all of the above neighbourhood properties---the
\CNP, \SCNP{} and \PIkNP{} for $k\ge1$, in particular---are well-defined
properties of {\em abstract\/} boundary sets.

Rather than catalogue a large list of topological properties in detail,
in this
section we shall just consider the situation where $T=SC$. Since it involves
hardly any more effort, we will seek to
generalise some of our results to the case where $T=\Pi_k$, $k\ge1$.

We begin by examining the issue of the simple connectedness of the
boundary sets themselves (\S\secref{SC bdy sets}),
then proceed to consider boundary sets which satisfy the \SCNP{}
(\S\secref{bdy sets sat'ing SCNP}).

\subsection{Simply connected boundary sets}
\seclabel{SC bdy sets}

We immediately note that,
just as in the case of connectedness, it may easily be seen
that simple connectedness does not pass to abstract boundary sets
(including abstract boundary points, i.e., members of the {\em a\/}-boundary).
We present here two examples of non-simply-connected boundary sets that are
equivalent to boundary points.  In the first example the boundary sets involved
are not isolated, whereas in the second example they are isolated.

\expandafter\ifx currentvolume\endcsname\relax 
	\else \newpage \fi

\begin{examples}\backtoroman
\item
	\label{ex:pt equiv nonisol circle}
	\begin{figure}
		\centering\leavevmode\epsfbox{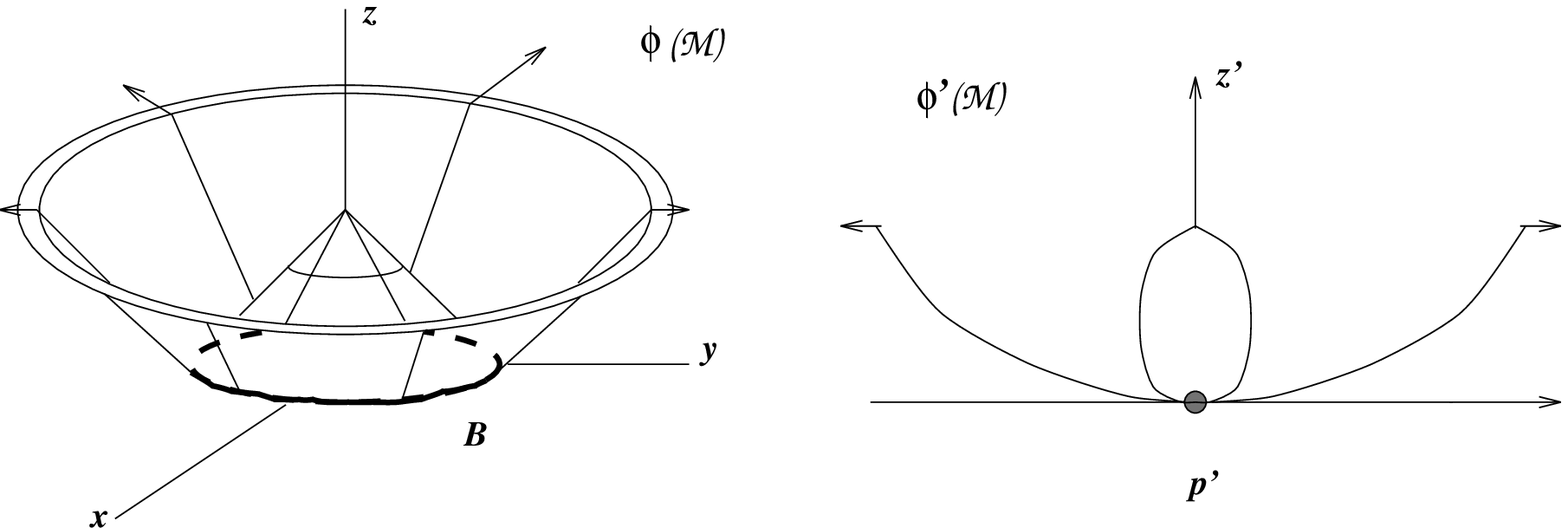}
		\caption{An illustration of
			Example~\protect\ref{ex:pt equiv nonisol circle}.
			On the left is depicted $\cM=\phi(\cM)\sub\Mhat$, and on the
			right is shown a cross-section of $\phi'(\cM)$, through a plane
			containing the $z'$-axis.
			The boundary sets $B$ and $\{p'\}$ are equivalent.
			\figlabel{pt equiv nonisol circle}}
	\end{figure}
	Let $f:\R^+\cup\{0\}\to \R$ (the domain is the non-negative reals)
	be the map
	$$
        x \mapsto \cases{
			1-x,    & for $0\le x\le 1$,\cr
			x-1,    & for $1<x<2$,\cr
			1,              & otherwise.\cr}
	$$
	Introduce cylindrical polar coordinates $r,\th,z$  on
	$\R^3\sm{}z$-axis.
	Let $\cM$ be the subset of
	$\R^3\sm{}z$-axis
	defined by
	$$
		\cM\defeq\{\: (r,\theta,z) \:\colon\: z > f(r) \:\}
	$$
	(see
	Figure~\figref{pt equiv nonisol circle}).
	Let $\Mhat=\Mhat'\defeq\R^3$, let $\phi$  be the inclusion, and let
	$$
        \phi':\cM\to\Mhat'\qquad\hbox{be given by}\qquad
		(r,\theta,z)\mapsto (r',\th',z')=(zr,\theta,z).
	$$
	This map is smooth and has Jacobian $z$, which vanishes nowhere on
	$\cM$.  It is thus immersive on $\cM$. By the Inverse Function Theorem
	(or by inspection), its inverse
	$ \phi'(\cM)\to\cM: (r',\th',z')\mapsto(r,\th,z)=(r'/z',\th',z')$ is also
	smooth.

	Let $B=\set (1,\th,0) for 0\le\th<2\pi\eset\sub\d_\phi $ and
	$p'=(0,0,0)\in\d_{\phi'} $.
	It is readily seen that $B\sim p'$,
	but $B$ is homeomorphic to $S^1$.

	This example  admits generalisation to higher dimensions,
	in the same manner as at the end of
	Example~\ref{ex:arb genus ex}.

	Thus, for any dimension $n\ge3$,
	a boundary point can be equivalent to a non-simply-connected boundary
	set.
	In particular, simple connectedness does not pass to the
	{\em a\/}-boundary.

\item\backtoroman
	\label{ex:15 from orig paper}
	Consider Example~15 of \citeasnoun{Scott94}.  There
	$\Mhat=\Mhat'=\R^n$, for some integer $n\ge 2$.  We
	introduce spherical polar coordinates $r,\range\th1{n-1}$, on the manifold
	$\cM\defeq\R^n\sm\{\vec0\}\approx\R^+\times S^{n-1}$.
	Let $\phi$ be the inclusion, and $\phi'$ be the map
	$$
		(r,\range\th1{n-1})\mapsto(r',\range{\th'}1{n-1})=(r+1,\range\th1{n-1}).
	$$
	Then let $B'\defeq \d_{\phi'}=\{r'=1\}\approx S^{n-1}$, and let $p$ be the
	origin $\vec0\in\Mhat$.
	Both $B'$ and $\{p\}$ are isolated boundary sets.

	We clearly have $B'\sim p$, yet for $n=2$, $B'\approx S^1$ is not
	simply connected.  (Indeed, $\pi_{n-1}(B')$ is non-trivial for any
	integer $n\ge2$.)

\end{examples}

\subsection{The simply connected neighbourhood property}
\seclabel{bdy sets sat'ing SCNP}

Although we have seen
(Corollary~\thmref{weak:pt conn},
Example~\ref{ex:disconn comp isol abs set not cont pt})
that, for $n\ge2$,
no {\em disconnected}, isolated boundary set is equivalent to
a boundary point, it is of interest to ask whether
some {\em connected}, isolated boundary sets have the
property that they are not equivalent to any boundary point.
We show here that there are indeed examples of this kind of boundary set.
In particular, we next give an example of a connected, isolated boundary set,
failing  to satisfy
the \SCNP, which is equivalent to no boundary point.  In fact,
for $n\ge3$,
an isolated boundary set which fails to
satisfy the \SCNP{}
is never equivalent to a boundary point.
(We shall delay the proof of this result until
Corollary~\thmref{isol, not SCNP equiv no pt}.)

\begin{example}\backtoroman
	\label{ex:conn comp isol abs set not cont pt}
	Let $\cM=\R^3\sm B$, where $B$ is a homeomorph
	of $S^1$: say $B=\{x^2+y^2=1,z=0\}$.
	Let $\Mhat=\R^3$ and $\phi$ be the inclusion, so that
	$\d_\phi=B$ is a non-simply connected, connected, isolated boundary
	set.

	We claim that $B$ does not satisfy the \SCNP.  Indeed,
	it is easy to see that any open neighbourhood $\cU$ of $B$ contains an
	open neighbourhood of the form
	$$
		\cW_\eta\defeq\bigcup_{b\in B}B_d(b,\eta),
	$$
	for some $0<\eta\le1$,
	where $d$ is the usual Euclidean metric on $\R^3$.
	For $0<\eta\le1$, the image of the loop
	$$
		\g_\eta:[0,1]\to\R^3:t\mapsto(1-\eta/2)(\cos2\pi t,\sin 2\pi t,0)
	$$
	is contained in $\cW_\eta\sm B=\cW_\eta\cap\phi(\cM)$.
	For any $0<\eta\le1$, the loop $\g_\eta$
	is clearly not homotopic to a constant map
	in $\cW_\eta\sm B=\cW_\eta\cap\phi(\cM)$; indeed, neither is $\g_\eta$
	homotopic to a constant map in $\cW_1\sm B=\cW_1\cap\phi(\cM)$.
	No open neighbourhood $\cV$ of $B$, contained in $\cW_1$, can
	be $SC$-nice: were it so, there would be some $0<\eta*<1$ for
	which $\cW_{\eta*}\sub\cV$, and the loop $\g_{\eta*}$ would be
	homotopic to a constant map in
	$\cV\sm B=\cV\cap\phi(\cM) \sub \cW_1\cap\phi(\cM)$.  This would be
	a contradiction, so that
	$\cW_1$ contains no $SC$-nice, open
	neighbourhoods of $B$. This means that $B$
	does not satisfy the \SCNP.

	Let $p'$ be an isolated boundary point of an envelopment
	$(\cM,\Mhat',\phi')$.
	We claim that $p'$ always satisfies the \SCNP.
	Fix a Riemannian metric $d'$ for $\Mhat'$, and
	define open  balls $B'_\e\defeq B_{d'}(p',\e)$, as in
	Lemma~\thmref{isol in sense that...}.
	It follows, by the same lemma, that for $\e$ sufficiently small we have
	$B'_{\e}\cap\phi'(\cM)= B'_{\e}\sm\{p'\}$, and that this set
	$B'_{\e}\cap\phi'(\cM)$ is homeomorphic to a ball $E^3$ without its centre.
	Hence $B'_{\e}\cap\phi'(\cM)$ is
	homotopic to a 2-sphere, and is thereby simply connected.
	It follows that $p'$ satisfies the \SCNP.

	By
	Theorem~\thmref{TNP invar},
	we have $B\not\sim p'$,
	where $p'$ is any isolated boundary point.
	Theorem~\thmref{isol invar}
	tells us, however, that
	any boundary point equivalent to $B$ must be isolated, so that
	$B$ cannot be equivalent to any boundary point; in other words,
	$[B]\nin\cB(\cM)$.
\end{example}

We know that a boundary set $B$ may be simply
connected, yet fail to satisfy the \CNP{}
(Example~\ref{conn, non-comp, SC, not CNP}
is an example of this, wherein  the set $B$ is both connected and
simply connected).  Such a set, then, cannot satisfy the \SCNP.
A simply connected boundary set may fail the \SCNP{} for more subtle
reasons than this, however---i.e., it is possible to find simply connected
boundary sets which satisfy the \CNP, yet which fail the \SCNP.
One example of this is the boundary point $p'$ of
Example~\ref{ex:pt equiv nonisol circle}.

We now complete our demonstration that the notions of simple connectedness
of boundary sets, on the one hand, and the \SCNP{} of boundary sets, on the
other, are logically independent.  To this end,
we will show in the next example that a boundary set $B$ may
satisfy the \SCNP, yet fail to be simply connected, itself.
This means we cannot find an analogue of
Theorem~\thmref{CNP is connected} for simple connectedness,
and demonstrates the difficulty of
finding criteria which ensure the invariance of simple connectedness
of boundary sets under boundary set equivalence.

\begin{figure}
	\maxpspicwid=.7\hsize
	\centering\leavevmode\epsfbox{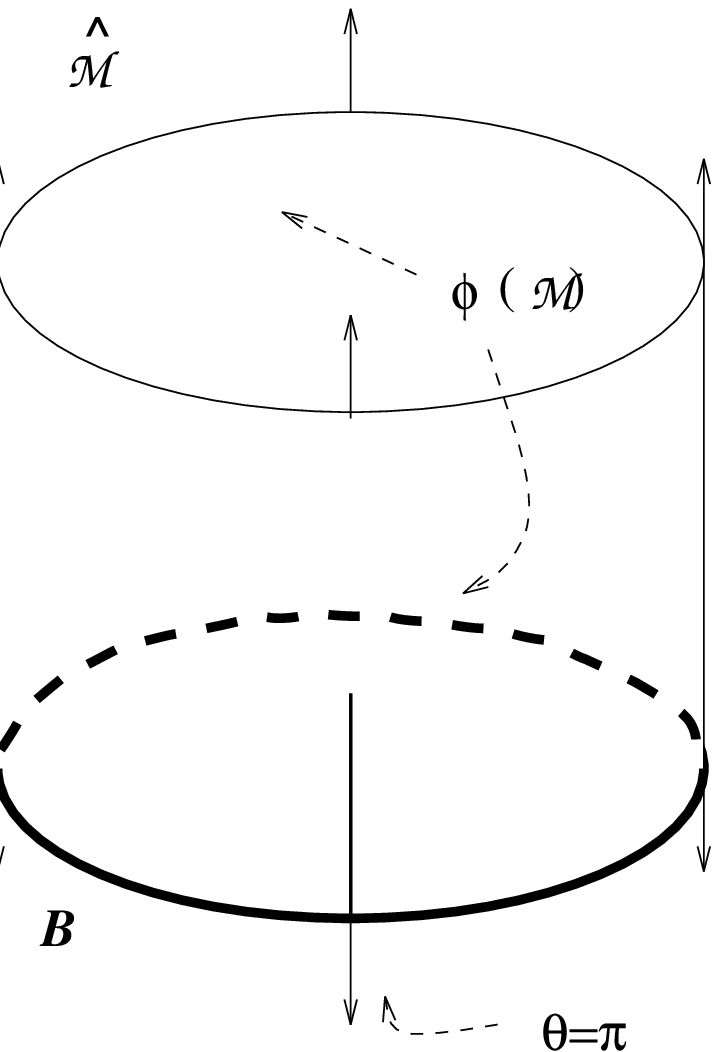}
	\caption{An illustration of
		Example~\protect\ref{SCNP but not SC}, in which
		$B$ satisfies the \protect\SCNP{} but is not simply connected.
		\figlabel{SCNP but not SC}}
\end{figure}
\begin{example}\backtoroman
	\label{SCNP but not SC}
	Let $\Mhat$ be a cylinder
	$$
		\Mhat\defeq \R\times S^1=\set (t,\th) for
		-\infty<t<\infty,
		-\pi<\th\le\pi
		\eset.
	$$
	Let $\cM\sub\Mhat$ be the region
	$$
		\set (t,\th)\in\Mhat for t>0, \th\ne\pi \hbox{ if } t\le1 \eset,
	$$
	and let $\phi$ be the inclusion---see
	Figure~\figref{SCNP but not SC}.
	So $\d_\phi=B\cup\set(t,\pi) for 0<t\le1\eset$, where $B$ is the circle
	$B\defeq \set (0,\th) for -\pi<\th\le\pi \eset$.
	It is easy to see that
	$B$, which is not simply connected, satisfies the \SCNP.

	This example  admits generalisation to higher dimensions,
	in the same manner as at the end of
	Example~\ref{ex:arb genus ex}, but here the manifold $X_0$ should be
	simply connected.
	Thus, for any dimension $n\ge2$, we have an example of a boundary set
	which is not simply connected, yet which satisfies the \SCNP.
	(In \S\secref{isol bdy subman}
	we shall show that, since $B$ is also a
	differentiable submanifold of
	$\Mhat$ admitting a nowhere-vanishing normal vector field,
	this pathology occurs {\em only\/} because $B$ fails to
	be isolated.)
\end{example}


\subsection{Isolated boundary points and the \protect\PIkNP}

When $n\ge3$, we can prove that there are no
{\em isolated\/} boundary points (and thus no {\em isolated\/}
abstract boundary points) which fail the \SCNP.
We can say considerably more than this, however.

\begin{theorem}
	\thmlabel{isol, not PIkNP equiv no pt}
	Let $k$ be a positive integer, and assume that the dimension $n$ of
	$\cM$ satisfies $n\ge k+2$.
	Then an isolated boundary point satisfies the \PIkNP.
	Thus, if  $n\ge k+2$,
	a boundary set which does not satisfy the \PIkNP{} is
	not equivalent to any isolated boundary point.
\end{theorem}
\begin{proof}
	Let $p$ be an isolated boundary point of the envelopment
	$(\cM,\Mhat,\phi)$,  and equip $\Mhat$ with a
	Riemannian metric $d$.  By
	Lemma~\thmref{isol in sense that...},
	for some $\e>0$, there is an open metric ball
	$B_\e\defeq B_d(p,\e)$ about $p$, of radius $\e$, which
	intersects $\d_{\phi}$ at (and only at) $p$,
	and is  such that $B_\e\sm\{p\}\sub\phi(\cM)$.
	Similarly, for all $\mu$ where
	$0<\mu<\e$, we let $B_\mu\defeq B_d(p,\mu)$.
	We have $B_\mu\cap\d_\phi=\{p\}$ and $B_\mu\sm\{p\}\sub\phi(\cM)$.
	Now each $B_\mu\sm\{p\}$ is homeomorphic to a ball without its
	centre, $E^n\sm\{\vec0\}$.  Thus each $B_\mu\sm\{p\}$ has the homotopy
	type of a sphere $S^{n-1}$.
	Since $\pi_k(S^{n-1})$ is trivial for
	$k<n-1$, it follows that
	$\{p\}$ satisfies the \PIkNP{} when $n\ge k+2$.
	This proves the first statement of the theorem.

	The second sentence is immediate.
	Assume that $B$ is an isolated boundary
	set which is equivalent to a boundary point $p'$.
	Then $p'$ is isolated, by
	Theorem~\thmref{isol invar},
	whence it satisfies the \PIkNP{}, for $n\ge k+2$.
	Thus $B$ also satisfies the \PIkNP, by
	Corollary~\thmref{TNP well-def}.
\end{proof}

Since \SCNP${}\equiv\Pi_1$NP, we have the following corollary as an example of
this theorem.

\begin{corollary}
	\thmlabel{isol, not SCNP equiv no pt}
	If $n\ge 3$, then any isolated boundary point satisfies the \SCNP.
	Thus, if $n\ge 3$,
	a boundary set which does not satisfy the \SCNP{} is
	not equivalent to any isolated boundary point.
\end{corollary}

\begin{remarks}
\begin{enumerate}
\item
	If $n=k+1$ then the result of
	Theorem~\thmref{isol, not PIkNP equiv no pt}
	is not true.  To see this,
	let $\Mhat=\R^{k+1}=\Mhat'$,
	and $\cM=\phi(\cM)=\R^{k+1}-\{p\}$ ($\phi={}$inclusion), where $p$
	is the origin $(0,\ldots,0)$.
	Let $\phi'$, in spherical polar coordinates, be the map
	$(r,\range\th1k)\mapsto(r+1,\range\th1k)$ as in
	Example~\ref{ex:15 from orig paper}.
	Then $p$, an isolated boundary point, is
	equivalent to a $k$-dimensional sphere $B'\approx S^k$.
	There is a base \seq{\cU'}A of neighbourhoods of $B'$ in $\Mhat'$,
	such that each $\cU'_A\cap\phi'(\cM)$ is homeomorphic to
	$S^k\times(0,1)$, and
	$$
		\pi_k(S^k\times(0,1))\approx\pi_k(S^k)\times\pi_k((0,1))\approx\Z.
	$$
	Hence no $\cU'_A\cap\phi'(\cM)$ is $k$-connected,
	and it follows that $B'$, and thus $p$, does not satisfy the \PIkNP.
\item
	The (integral) homology groups of the sphere $S^{n-1}$ are
	$$
		H_i(S^{n-1})\approx\cases{
			0,& if $i\ne0,n-1$,\cr
			\Z, & if $i=0$ or $i=n-1$,\cr}
		\qquad 0\le i\le n-1.
	$$
	Upon examining the proof of
	Theorem~\thmref{isol, not PIkNP equiv no pt},
	it is easily seen that one can replace the \PIkNP{} with
	the \HOMkNP{} in the statement of that theorem, to obtain a new
	result.
	To do this, we need conditions on
	$k$ and $n$ that will ensure the triviality of $H_k(S^{n-1})$, and
	clearly the condition $1\le k\le n-2$ suffices.  Thus we have the
	following result.

	\begin{theorem}
		Let $k$ be a positive integer, satisfying $1\le k\le n-2$, where
		$n$ is the dimension of $\cM$.
		Then an isolated boundary point satisfies the \HOMkNP.
		Thus, if $1\le k\le n-2$,
		a boundary set which does not satisfy the \HOMkNP{} is
		not equivalent to any isolated boundary point.
	\end{theorem}
\end{enumerate}
\end{remarks}
\section{An application to isolated boundary submanifolds}
\seclabel{isol bdy subman}

Note that
Theorem~\thmref{isol, not PIkNP equiv no pt}
and
Corollary~\thmref{isol, not SCNP equiv no pt}
do not directly help us in deciding whether a boundary set $B$,
which fails to be simply connected {\bf\em itself}
(or is not $k$-connected, for some $k>1$),
can be equivalent to an {\em isolated\/} boundary point.
In this section we seek conditions on $B$ that will allow us to
answer such a question.

\begin{definition}\backtoroman
	A boundary set $B$ of an envelopment $(\cM,\Mhat,\phi)$, which is also a
	differentiable submanifold of $\Mhat$, will be called a {\em boundary
	submanifold}.  If $B$ is also of codimension~1 in $\Mhat$, it will
	be called a {\em boundary hypersurface}.
\end{definition}

We will show that isolated boundary submanifolds $B$,
which fail to be $k$-connected (i.e., which have non-trivial $k$th
homotopy group), and also obey a fairly easily checked criterion,
fail to satisfy the \PIkNP. Thus, such a $B$ cannot be
equivalent to a boundary set satisfying the \PIkNP. In particular,
if $n\ge k+2$,
$B$ cannot be equivalent to a (necessarily isolated) boundary point.

In considering an isolated boundary submanifold $B$ of codimension
$\ell>1$,
one might hope to find open neighbourhoods of $B$
homeomorphic to a product $B\times E^\ell$, with $B$ carried onto
$B\times\{\vec0\}$.
Unfortunately, such neighbourhoods do not exist in general.
The problem is to do with the possible non-triviality of the normal
bundle of a submanifold.%
\footnote{
	If this is non-trivial, of course, the submanifold
	cannot be contractible ({\em in itself}, that is), by
	Corollary~4.2.5 of \citeasnoun{Hirsch76}---every vector bundle over a
	contractible, paracompact base space is trivial.
}
To deal with this we use more general
neighbourhoods called ``tubular neighbourhoods''.
(In Appendix~\secref{bundles},
we give a few definitions that may be of use to those unfamiliar with
some of the following terminology.)

\begin{definition}\backtoroman
	Let $\cV$ be a smooth manifold, and
	let $\cM\sub\cV$ be a differentiable submanifold. A {\em tubular neighbourhood
of $\cM$
	{\rm(or} for $(\cV,\cM)$}) is a pair $(f,\xi)$, where $\xi=(p,E,\cM)$
	is a vector bundle over $\cM$ and $f:E\to \cV$ is an embedding such that
	\begin{enumerate}\thisenumdot{arabic}
	\item
		$f|_\cM={}$the identity on $\cM$, where $\cM$ is identified with the
		zero section of $\xi$;
	\item
		$f(E)$ is an open neighbourhood of $\cM$ in $\cV$.
	\end{enumerate}
	We often simply refer to $\cW=f(E)$ as a tubular neighbourhood of $\cM$.
	To $\cW$ is associated a retraction
	$q:\cW\to \cM$, making $(q,\cW,\cM)$ a vector bundle, isomorphic to $\xi$,
	whose zero section is the inclusion $\cM\sub \cV$.%
	\footnote{
		Remember, a retraction is a left inverse to the inclusion in the
		topological category.
	}

	When $\xi$ is the normal bundle of $\cM$ in $\cV$, $\cW$ is called a
	{\em normal tubular neighbourhood}.
\end{definition}

Note that differentiability of all
manifolds is required in the above definition,
since one requires $f$ to be an embedding.

We now state the very powerful {\em Tubular Neighbourhood Theorem.}

\begin{theorem}[{\citeasnoun[Theorem~4.5.2]{Hirsch76}}]
	\thmlabel{TNT}
	Any differentiable submanifold $\cN$ of a differentiable manifold $\cM$ has
	a tubular neighbourhood $(f,\xi)$ \rom(or $\cW$\rom).
	Furthermore, $(f,\xi)$ can be chosen so that $\xi$ is the normal
	bundle of $\cN$, with respect to any Riemannian metric on $\cM$.
\end{theorem}

In fact, by ``retracting along the fibres'' of any tubular
neighbourhood and, in particular, any normal tubular
neighbourhood, we see that the collection of
normal tubular neighbourhoods of $\cN$ forms a base for the neighbourhood
system
of $\cN$.

Using normal tubular neighbourhoods, we can relate the condition of
$k$-connectedness
of an {\em isolated\/} boundary submanifold $B$,
to the \PIkNP{} for $B$.

\begin{theorem}
	\thmlabel{PIk(B) nontriv => fails PIkNP}
	Let $B$ be a connected,  isolated boundary submanifold
	of the envelopment $(\cM,\Mhat,\phi)$.
	Fix any Riemannian structure on $\Mhat$, with respect to which the
	normal bundle $NB$ of $B$ is defined.
	Assume that there is a section of the normal bundle $NB$ of $B$ in
	$\Mhat$ which does not vanish at any point of $B$.
	If $B$ is not $k$-connected, then $B$ does not satisfy the \PIkNP.
\end{theorem}
\begin{proof}
	Let $\s:B\to NB$ be a nowhere-vanishing section of $NB$.
	Since $B$ is isolated, there is a neighbourhood $\cU$ of $B$ in
	$\Mhat$ such that $\cU\cap\d_\phi=B$.
	Let $\cW$ be a normal tubular neighbourhood of $B$ contained in $\cU$
	(c.f.\ %
	Theorem~\thmref{TNT}),
	with projection
	$q:\cW\to B$,  and let $f:NB\to\cW$ be an embedding, as in the
	definition of tubular neighbourhoods.
	We will use the map
	$f\circ\s$ to ``lift'' a map $\l:S^k\to B$ to a map
	$S^k\to\cW\cap\phi(\cM)$. Before doing this, though, we must check
	that $\s$ ``points into $\phi(\cM)$'', and replace it if this is not
	the case.

	First, let us assume that the codimension of $B$
	in $\Mhat$ is greater than one.
	Now each point $p$ of $B$
	has a neighbourhood $X_p\sub\cW$ homeomorphic to $\R^n$, with $B\cap X_p$
	corresponding to a linear subspace $V\sub\R^n$ of dimension $m\le n-2$.%
	\footnote{%
		We use the Frobenius Theorem here.
	}
	We claim that
	$\R^n\sm V$ is connected. Non-singular linear transformations of
	$\R^n$ are homeomorphisms, whence we may assume that $V=\{\vec x=(\range
	x1n)\in\R^n \::\: x_{m+1}=\cdots=x_n=0\}$. The claim is then
	obvious,
	and thus $X_p\sm B \approx \R^n\sm V$ is connected.
	From this it follows that $\bigcup_{p\in B}(X_p\sm B)$, being the union of
	connected, open sets, none of which is disjoint from all the others
	(since $B$ is connected),
	is also connected, whereby it is contained in $\phi(\cM)$
	(c.f.\ the proof of
	Lemma~\thmref{isol in sense that...}).
	Finally, we ``normalise'' $\s$ in the following manner.
	We can easily construct a smooth $t:B\to\R^+$ such that
	$f\circ[t(p)\cdot \s(p)]\in X_p\sm B$,
	for all $p\in B$
	(recall that $\s$ is nowhere-vanishing).
	Then the section $t\cdot\s$ (defined as the map
	$p\mapsto t(p)\cdot\s(p)$) satisfies
	$f\circ(t\cdot\s)\circ\l(S^k)\sub\cW\cap\phi(\cM)$.
	We replace $\s$ with $t\cdot\s$.

	It remains to consider the case where
	the codimension of $B$ in $\Mhat$ is precisely one,
	i.e., where $B$ is  a boundary hypersurface.
	In a similar manner to the argument of the previous paragraph, we note that
	each point $p$ of $B$
	has a neighbourhood $X_p\sub\cW$ homeomorphic to $\R^2$, with $B\cap X_p$
	corresponding to an axis $\R$.
	We can also choose the homeomorphisms so
	that $X_p\cap\phi(\cM)$ either corresponds to $\R^2\sm x$-axis, or
	to the (open) upper half-plane.

	As before, we can ``normalise'' $\s$ so that
	$f\circ\s(B)\sub \bigcup_{p\in B}(X_p\sm B)$.
	Now at {\em any\/} point $p\in B$, either
	$f\circ\s(p)\in X_{p}\cap\phi(\cM)$ or
	$f\circ(-\s)(p)\in X_{p}\cap\phi(\cM)$
	(or both are true),
	where $-\s\defeq-1\cdot\s$.
	If only the latter option is true,
	then we replace $\s$ with $-\s$
	(this corresponds to
	swapping the ``inward'' or ``outward'' character of the ``normal field''
	$\s$).  It now follows, since the image
	$f\circ\s\circ\l(S^k)$ is a connected subset of $\cW$, and since
	$f\circ\s(p)\in\phi(\cM)$ for any $p\in B$ (and, in particular, for
	$p\in\Im \l$),
	that $f\circ\s\circ\l(S^k)\sub\cW\cap\phi(\cM)$.

	In both these cases, then, we may assume that
	$f\circ\s\circ\l(S^k)\sub\cW\cap\phi(\cM)$ (i.e., this assumption is
	valid, regardless of the codimension of $B$ in $\Mhat$).

	Since
	$\pi_k(B)$ is non-trivial, there is a map $\l:S^k\to B$ which is not
	homotopic to a constant map.
	Consider the map $\l':S^k\to \cW\cap\phi(\cM)$ defined by
	$$
		\l' \defeq f\circ\s\circ\l.
	$$
	Were $\l'$ homotopic to a constant map in $\cW\cap\phi(\cM)$,
	such a homotopy
	$$
		F:S^k\times I\to\cW\cap\phi(\cM),
		\qquad F(\spdot,0)=\l'(\spdot),
		\qquad F(\spdot,1)=w_0\in\cW\cap\phi(\cM),
	$$
	would give rise to a homotopy
	$$
		q\circ F:S^k\times I\to B,
		\qquad q\circ F(\spdot,0)=\l(\spdot),
		\qquad q\circ F(\spdot,1)=q(w_0).
	$$
	This would be a contradiction, since we assumed
	that $\l$ was not homotopic to a constant map.
	Thus the ``lift'' $\l'$ cannot be homotopic to a
	constant map in $\cW\cap\phi(\cM)$.

	We have now established most of the machinery necessary for the proof.
	To proceed, let us assume that $B$ satisfies the \PIkNP, and attempt to
	derive a contradiction.  Let $\cW_0$ be a normal tubular
	neighbourhood of $B$ as in the above paragraphs (with $q_0:\cW_0\to B$ the
	projection, and $f_0:NB\to\cW_0$ an embedding, as in the definition
	of tubular neighbourhoods).
	Consider the ``lift'' $\l'_0\defeq f_0\circ\s\circ\l$.
	Since $B$ satisfies the \PIkNP,
	$\cW_0$ contains an open $\Pi_k$-nice neighbourhood $\cV$ of $B$.
	Now $\cV$ contains a normal tubular neighbourhood $\cW_1$ of $B$,
	with projection $q_1:\cW_1\to B$, and with embedding
	$f_1:NB\to\cW_1$.
	We may assume that $\cW_1$ is obtained by ``shrinking'' $\cW_0$:
	to be more specific, $f_0\circ{f_1}^{-1}:\cW_1\to\cW_0$ restricts on
	each fibre (which has the structure of a vector space)
	to be a dilatation---see
	Figure~\figref{lifts}.
	\begin{figure}
		\maxpspicwid=.8\hsize
		\centering\leavevmode\epsfbox{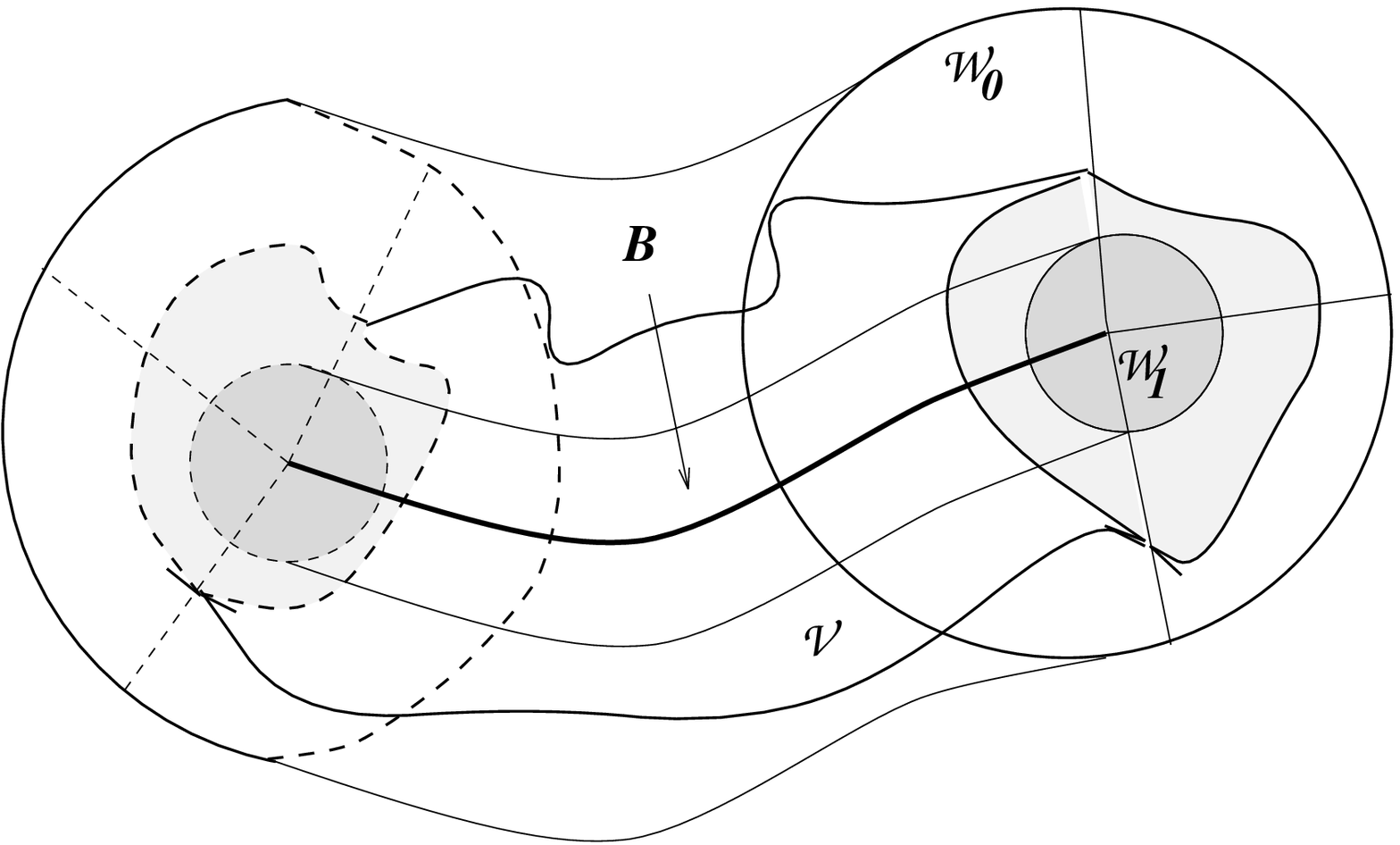}
		\caption{An illustration for the proof of
			Theorem~\protect\thmref{PIk(B) nontriv => fails PIkNP}.
			Here $\cV$ is $\Pi_k$-nice, $\cW_0$ and $\cW_1$ are normal tubular
			neighbourhoods of $B$, and $\cW_1\sub\cV\sub\cW_0$.
			\figlabel{lifts}}
	\end{figure}

	Consider $\l'_1\defeq f_1\circ\s\circ\l$, which
	satisfies $\l_1'(S^k)\sub\cV\cap\phi(\cM)$.
	By our assumption on the relationship between $\cW_1$ and $\cW_0$
	at the end of the previous paragraph, we have that
	$\l_1'$ and $\l_0'$ are homotopic
	in $\cW_0$. Hence $\l_1'$ is not homotopic to a constant map in
	$\cW_0\cap\phi(\cM)$, and it follows that it cannot
	be so in $\cV\cap\phi(\cM)$.
	This is a contradiction because we assumed that $\cV$ is
	$\Pi_k$-nice, and
	the result is established.
\end{proof}

In the above proof,
the condition that $NB$ have a nowhere-vanishing section is a convenient
tool for ``lifting'' a map $S^k\to B$ to a map from $S^k$ to
$\cW\cap\phi(\cM)$, where $\cW$ is a normal tubular neighbourhood of $B$.
This condition should be easily verified (if
true), for specific examples---%
especially when $B$ is contained in a single
coordinate chart of $\Mhat$.
One such example is
Example~\ref{ex:conn comp isol abs set not cont pt}, where $NB$ clearly
has a nowhere-vanishing section.
We thereby see that
Theorem~\thmref{PIk(B) nontriv => fails PIkNP}
is a significant generalisation of
Example~\ref{ex:conn comp isol abs set not cont pt}.

The aforementioned
condition can easily be seen to be satisfied {\em automatically\/}
in certain cases, two of which now follow.

\begin{corollary}
	\thmlabel{half-codim PIk(B) nontriv -> fails PIkNP}
	Assume that $\dim\cM=2n$, where $n$ is an odd integer.
	Let $B$ be an oriented, connected, isolated boundary submanifold
	of the envelopment $(\cM,\Mhat,\phi)$, where $\dim B=n$
	and $\Mhat$ is oriented.
	Then $B$ does not satisfy the \PIkNP,
	if $B$ is not $k$-connected.
\end{corollary}
\begin{proof}
	Fix any Riemannian structure on $\Mhat$, with respect to which the
	normal bundle $NB$ of $B$ is defined.
	Then, the result follows immediately from
	Theorem~\thmref{PIk(B) nontriv => fails PIkNP},
	once we note that a necessary and sufficient condition for $NB$ to
	have a nowhere-vanishing section is that the topological Euler
	characteristic $X(NB)$ of the normal bundle vanishes.
	(See \S5.2 of \citeasnoun{Hirsch76} and,
	in particular, Theorems~5.2.2 and~5.2.10.)
	This invariant does indeed vanish for $n$-dimensional oriented
	vector bundles
	over compact, oriented $n$-manifolds
	\cite[Theorem~5.2.5(b)]{Hirsch76}
	and, since $NB$ inherits the orientation of $\Mhat$ by  the Tubular
	Neighbourhood Theorem
	(Theorem~\thmref{TNT}),
	the result is proven.
\end{proof}

\begin{remark}
	It is somewhat unfortunate that we cannot use the
	simple argument given above for the cases where
	$B$ is of {\em arbitrary\/}
	dimension $1\le m\le 2n-1$.
	As noted above, though, the assumption of
	Theorem~\thmref{PIk(B) nontriv => fails PIkNP},
	that $NB$ has a nowhere-vanishing section, was merely a tool for
	obtaining ``lifts'' of maps $S^k\to B$.  Even though this assumption
	will fail for some compact boundary submanifolds (though certainly not for
	all such; e.g., see
	Example~\ref{ex:conn comp isol abs set not cont pt}),
	it could be possible to obtain ``lifts'' that are
	still suitable for this purpose, by a different method.%
	\footnote{%
		One example of a situation where the normal bundle of a compact
		submanifold has no nowhere-vanishing section,
		is given  by letting $\Mhat$ be a M\"obius band, and
		letting $B$ be the ``central circle'' of $\Mhat$.
		We note that, although $B$ is a non-simply connected, isolated
		boundary 1-submanifold of a 2-manifold, $\Mhat$ is
		not oriented, so the above proof fails.
		The authors are unaware of any
		instances of connected, {\em orientable},
		non-$k$-connected, isolated boundary
		submanifolds $B$ of an oriented manifold $\Mhat$,
		satisfying the \PIkNP{} (where $\dim B\ne\half \dim\Mhat$).
	}
\end{remark}

If $\Mhat$ is simply connected and $B$ is a boundary hypersurface
(i.e., it has codimension~$1$), then we can obtain a result similar to
Corollary~\thmref{half-codim PIk(B) nontriv -> fails PIkNP},
without the condition of
orientability of $B$ or $\Mhat$.

\begin{corollary}
	\thmlabel{1-codim PIk(B) nontriv -> fails PIkNP}
	Let $B$ be a connected, isolated boundary hypersurface of a simply connected
	enveloping manifold $\Mhat$.
	If $B$ is not $k$-connected,
	then $B$ fails to satisfy the \PIkNP.
\end{corollary}
\begin{proof}
	Assume that the conditions of the statement hold.
	Fix any Riemannian structure on $\Mhat$, with respect to which the
	normal bundle $NB$ of $B$ is defined.
	By
	Theorem ~4.4.6
	and
	Lemma~4.4.4
	of \citeasnoun{Hirsch76}, a connected, compact, boundary hypersurface
	of a {\em simply connected\/} manifold $\Mhat$ has trivial normal bundle, so
	$NB$ is trivial.  Thus $NB$ admits a nowhere-vanishing section, and
	Theorem~\thmref{PIk(B) nontriv => fails PIkNP}
	gives the result.
\end{proof}

We now apply these corollaries to the abstract boundary in the usual way.

\begin{corollary}
	Let $k$ be  a positive integer,
	and let $B$ be a connected,
	isolated boundary submanifold
	of an envelopment $(\cM,\Mhat,\phi)$.
	Assume that either
	\begin{itemize}
	\item
		$n$ is an odd integer, $\dim\cM =\dim\Mhat=2n\ge k+2$,
		$\dim B=n$, and both $\Mhat$ and $B$ are oriented, or
	\item
		$B$ is a boundary hypersurface of the simply connected
		manifold $\Mhat$, and $\dim\cM=\dim\Mhat=n\ge k+2$.
	\end{itemize}
	If $B$ is not $k$-connected, then $B$ cannot be equivalent to any
	boundary point.
\end{corollary}
\begin{proof}
	This follows immediately from
	Corollary~\thmref{half-codim PIk(B) nontriv -> fails PIkNP},
	Corollary~\thmref{1-codim PIk(B) nontriv -> fails PIkNP},
	and
	Theorem~\thmref{isol, not PIkNP equiv no pt}.
\end{proof}

\section{Summary and discussion}

The {\em a\/}-boundary construction
deals with {\em arbitrary\/} smooth embeddings  of a manifold into
other manifolds of the same dimension.
In this paper we have demonstrated that,
even when dealing with purely topological properties of
boundary sets such as compactness, connectedness and simple
connectedness, much can be said about the ways in which these properties
are preserved (or otherwise) under boundary set equivalence.
Indeed, it was shown
that compactness is invariant under boundary set equivalence.  We
proceeded to define a property, ``isolation'',
of boundary sets, which encapsulates the notion of a
boundary set being ``distant'' from other points of the boundary (in the
same envelopment).
Isolation is also a property which is
invariant under boundary set equivalence.

We introduced the wide-ranging concept of
``neighbourhood properties'' of boundary sets, examples of which include the
``connected neighbourhood property'' (\CNP)
and the
``simply connected neighbourhood property'' (\SCNP).
The utility of these
neighbourhood properties
lies in the fact that they are all
invariant under boundary set equivalence.
In addition, we proved that a boundary set satisfying the \CNP{}
is always, itself, connected.

The \SCNP{} and, more generally, the
``$k$-connected neighbourhood property'' (\PIkNP),
were examined in some detail.
It was shown that neither of these properties of boundary sets $B$
necessarily imply the appropriate conditions on the homotopy groups of $B$
itself (i.e., triviality of $\pi_1(B)$ and $\pi_k(B)$,  respectively).
Both isolated and non-isolated boundary sets provided valuable examples
here.
Finally, the special case of connected, isolated
boundary submanifolds was considered.  It was shown that, under certain
circumstances,
if $B$ is such a submanifold which is not $k$-connected, then it cannot
satisfy the \PIkNP.
It was noted that,
if $\dim\cM\ge k+2$, this means, in particular, that
$B$ cannot be a representative  boundary set of an abstract
boundary point.

Thus, it is quite tenable to study the equivalence properties of boundary sets
before one even considers the classification of boundary points
(and of abstract boundary points)
into regular boundary points, points at infinity, singularities, {\em etc}.
The problems one encounters when entering the realm of this
classification, not surprisingly,
tend to be inherently geometrical rather than topological in nature.
Such matters are the subject of on-going research, and will be considered
in future papers.

\section*{Acknowledgements}

The authors wish to thank Marcus~Kriele for the suggestion that led,
originally, to
Example~\ref{ex:conn comp isol abs set not cont pt}.
One of the authors (SMS) would like to express her appreciation to
	the Institute for Theoretical Physics, University of California,
	Santa Barbara,
where part of this work was undertaken, during a visit in 1993.

\appendix
\section{Submanifolds and normal bundles}
\seclabel{bundles}

Here, we summarise the concepts from differential topology which are used in
\S\secref{isol bdy subman}.
We recommend
the book by Hirsch \cite{Hirsch76} (from which much of the following
comes {\em verbatim\/}), although many
books on differential topology contain a treatment of these
concepts.

\begin{definitions}\backtoroman
	Let $p:E\to B$ be a continuous map between topological spaces.
	A {\em vector bundle chart\/} $(\vp,\cU)$
	on $(p,E,B)$ with domain $\cU$ and dimension $n$ is a homeomorphism
	$\vp:\inv p(\cU)\approx \cU\times\R^n$ ($\cU\sub B$ open)
	\expandafter\ifx\csname diagram\endcsname\relax
	such that
	$\vp\circ\inv p(x)=\{x\}\times\R^n$ for any $x\in\cU$.
	\else
	which makes the following
	commute:
	$$
		\diagram
			\inv p(\cU) \rrto^\vp \drto_p && \cU\times\R^n \dlto^{\pi_1} \\
			& \cU\\
		\enddiagram
		\qquad\hbox{where } \pi_1(x,v)=x.
	$$
	\fi
	For each $x\in B$ the  composition
	$$
		\inv p(x) \stackrel\vp\longrightarrow \{x\}\times\R^n
		\longrightarrow\R^n
	$$
	will be denoted $\vp_x$.

	A {\em vector bundle atlas\/} $\Phi$ is a family of
	vector bundle charts with values in the same $\R^n$, whose domains
	cover $B$, such that whenever $(\vp,\cU),(\psi,\cV)\in\Phi$
	and $x\in \cU\cap \cV$,
	the homeomorphism $\psi_x\inv{\vp_x}:\R^n\to\R^n$ is linear;
	furthermore the map
	$$
		\cU\cap \cV\to\GL(n,\R):x\mapsto \psi_x\inv{\vp_x}
	$$
	is required to be continuous.

	If $\xi=(p,E,B)$ admits a vector bundle atlas, it is
	a {\em vector bundle\/} of ({\em fibre\/}) {\em dimension\/} $n$ with
	{\em projection\/} $p$,
	{\em total space\/} $E$
	and {\em base space\/} $B$.  $E_x\defeq\inv p(x)\approx\R^n$
	is called the {\em fibre\/} over $x$.
	$\xi$ may be called an {\em $n$-plane bundle}, and denoted $E\to B$
	or even $E$.

	A {\em section\/} of $\xi$ is a map $\s:B\to E$ with $p\circ
	\s={}$identity: thus $\s(x)\in E_x$.
	The {\em zero section\/} of $\xi$ is the map $Z:B\to E$ with $Z(b)$
	being the zero element of the fibre $E_x$, and we can identify $B$
	with $Z(B)\sub E$.

	A map $F:E_0\to E_1$ between vector bundles $\xi_i=(p_i,E_i,B_i)$,
	$i=0,1$, is a {\em fibre map\/} covering $f:B_0\to B_1$ if
	\expandafter\ifx\csname diagram\endcsname\relax
	$ p_1\circ F = f\circ p_0$.
	\else
	$$
		\diagram
			E_0 \rrto^F \dto_{p_0} && E_1 \dto^{p_1} \\
			B_0 \rrto_f && B_1\\
		\enddiagram
	$$
	commutes.
	\fi
	If, as well, each $F_x\defeq F|_{E_{0x}}:E_{0x}\to (E_1)_{f(x)}$
	is linear, $F$ is a {\em morphism\/} of vector bundles.  If each $F_x$ is
	$\left\{\begin{array}{@{}l@{}}
		\hbox{injective}\\
		\hbox{surjective,}\\
		\hbox{bijective}\\
		\end{array}\right.$
	we call $F$ a
	{\em
	$\left\{\begin{array}{@{}l@{}}
		\hbox{monomorphism}\\
		\hbox{epimorphism,}\\
		\hbox{bimorphism}\\
		\end{array}\right.$}
	respectively.
	A bimorphism covering a homeomorphism is an {\em equivalence}, and
	if $B_0=B_1$, a bimorphism covering the identity map is an {\em
	isomorphism\/}: $\xi_0\approx \xi_1$.

	The above definitions can be phrased ``in the $C^r$ category''
	(i.e., when all topological spaces and
	maps are required to be $C^r$ manifolds and maps,
	respectively), and $r=\infty$
	is
	allowed.  In this manner, one defines a {\em $C^r$ vector bundle},
	$0\le r\le\infty$.
	When $r\ge1$, the zero section identifies $B$ with a differentiable
	submanifold of $E$.
\end{definitions}

Barring analyticity,
the degree of differentiability of vector bundles
(over smooth manifolds)
is not really important, because of the following theorem.
(Compatibility of vector bundle structures is defined similarly to that
for differential structures on manifolds.)

\begin{theorem}[{\citeasnoun[Theorem 4.3.5]{Hirsch76}}]
	Every $C^r$ vector bundle $\xi$, $0\le r\le\infty$,
	over a $C^\infty$ manifold $ \cM $
	has a compatible $C^\infty$ vector bundle structure, unique up to
	$C^\infty$ isomorphism.
\end{theorem}

This is quite a deep result, using at its heart transversality (the
Morse-Sard Theorem) \cite[\S3]{Hirsch76}.  It is quite surprising for
the case $r=0$.

Finally, we define the normal bundle of a submanifold, and orientability
of manifolds.

\begin{definitions}\backtoroman
	\label{def:normal bundle}
	Let $\cN$ be a differentiable manifold with
	a Riemannian metric (or at least an {\em orthogonal structure\/}
	\cite[\S4.2]{Hirsch76}),
	and let $\cM\sub \cN$ be a differentiable submanifold.
	Let $T_\cM\cN$ denote
	the bundle of vectors of $T\cN$ at points of $\cM$.
	Since $T\cM$ can be viewed as a subset of $T_\cM\cN$ in a natural
	way, it makes sense to consider
	the bundle, over $\cM$,  whose fibre at $x\in\cM$ is the
	orthogonal complement of $T_x\cM\sub T\cM$ in $(T_\cM\cN)_x$.
	This bundle
	is called the ({\em geometric\/}) {\em normal bundle\/} of $\cM$ in
	$\cN$ and is often denoted $N\cM$, or $T^\perp\cM$.

	A manifold $\cM$ is called {\em orientable\/} if it admits
	an atlas of coordinate charts
	$$
		\{(U,\vp),(V,\psi),\ldots\},
	$$
	the transition functions
	$\vp\circ\psi^{-1}:\psi(U\cap V)\to \vp(U\cap V)$ of which have
	Jacobians of the same sign at each point.  (There are several other
	equivalent definitions of orientability, but this is one of the
	easiest to state.)
\end{definitions}



\end{document}